\documentstyle[prb,preprint,aps]{revtex}
\newcommand{\be}{\begin{equation}}
\newcommand{\ee}{\end{equation}}
\newcommand{\bea}{\begin{eqnarray}}
\newcommand{\eea}{\end{eqnarray}}
\newcommand{\OP}{\vec{\psi}}
\newcommand{\SG}{\vec{\sigma}}
\newcommand{\M}{\vec{m}}
\newcommand{\U}{\vec{u}}
\newcommand{\F}{{\cal F}}
\newcommand{\A}{\omega_{0}}
\newcommand{\B}{\omega_{2}}
\newcommand{\C}{\omega_{1}}
\begin{document}

\title{Vortex Velocity Pair Correlations}
\author{Gene F. Mazenko }
\address{The James Franck Institute and the Department of Physics\\
The University of Chicago\\
Chicago, Illinois 60637}
\date{\today}
\maketitle
%

%
%  ABSTRACT
%
\begin{abstract}

The vortex velocity probability distribution for two distinct vortices is
determined for the case of phase-ordering kinetics in systems with
point defects. The n-vector model driven by  time-dependent
Ginzburg-Landau dynamics for a nonconserved order parameter is considered.
The
description includes
the effects of other vortices and order parameter fluctuations.
At short distances the most probable configuration is that a
vortex-antivortex pair
have only a nonzero relative velocity which is
inversely proportional
to the distance between them.  The coefficient of proportionality
is determined explicitly.

\end{abstract}


\section{Introduction}

It seems plausible that much of the structure one sees in the phase
ordering of many materials\cite{1,2}
can be associated with the evolution and
correlation among defects\cite{MERMIN79} like vortices, monopoles,
disclinations, etc.
These topologically robust objects grow out of the
frustration suffered by a system with a continuous symmetry which is
thermodynamically driven to align in a broken symmetry state.
In the case of the n-vector model with the number
of components ($n$) of the order parameter
equal to the spatial dimensionality
($d$) one has point defects which are
vortices for $n=2$ and monopoles for $n=3$.
Because of
the conservation of topological charge, the ordering in these
systems is through the
charge conserving process of
vortex-antivortex annihilation. Topological constraints render the
ordering in such systems to be
largely independent of the microscopic details
of the material.  In this paper the following question
is addressed:  What is
the probability, given a vortex at position $\vec{r}_1$ with velocity
$\vec{v}_{1}$, that one will find a vortex at position $\vec{r}_{2}$
with velocity
$\vec{v}_{2}$? Clearly in answering this question,
we obtain a
tremendous amount of information about the dynamics of vortices.

The calculation of the two vortex velocity probability distribution
is a very involved process.  In principle one could probe vortex
dynamics by applying a force.  Unfortunately in neutral systems it
is very difficult to couple directly to the vortices.  The two
vortex velocity probability distribution serves this purpose by looking
at the motion of one vortex in the fixed presence of another vortex a
known distance away.

The  physical results of this calculation, carried out in detail
for $n=d=2$, are relatively simple to state.
The appropriate probability distribution is a function only of the scaled
velocities $\vec{u}_{i}¥=\vec{v}_{i}¥/\bar{v}$ for $i=1$ or $2$, and
the scaled separation $\vec{x}=(\vec{r}_{1}-\vec{r}_{2})/L(t)$.
Here $L(t)$ is the characteristic length in the problem which grows
with time $t$ after a quench
as $t^{1/2}$ in the present case and drives the scaling behavior\cite{1}
found in the problem.  The characteristic velocity
$\bar{v}$, defined carefully below,  is inversely proportional to
$L(t)$.
For a given scaled separation $x$ between two chosen vortices,
the most probable configuration
corresponds, as expected, to a state with zero total momentum and
a nonzero
relative momentum only along the axis connecting the vortices:
\be
\vec{v}_{1}=-\vec{v}_{2}=v\hat{x}  ~~~.
\ee
Moreover there is
a definite nonzero value for $v=v_{max}$ for a given value of $x$.  These
most probable values are given as a function of $x$ in Fig.1.  The most
striking feature of these results is that for small $x$ the
most probable velocity
goes as
\be
v_{max}=\frac{\kappa}{R}
\ee
where $R$ is the unscaled separation between the vortices and
$\kappa =2.19$ in dimensionless units defined below.
The result giving $v_{max}$ inversely proportional to $R$ is consistent with
overdamped dynamics where the relative velocity of the two vortices
is proportional to the
force which in turn is the derivative of a potential which is logarithmic
in the separation distance.
Thus these most probable results are consistent with the short-
distance behavior being dominated by the annihilation of vortex-
antivortex pairs.
{}From previous work\cite{MWann} we know that
there is low probability of finding like signed vortices at short
distances.  Thus our results giving the velocity as a function
of separation should be interpreted in terms of annihilating
vortex-antivortex pairs.  The results for same signed vortices
can also be carried out but is considerably more involved as
discussed below.

The work here builds on the work in Ref.\onlinecite{MAZENKO97}
where the single vortex velocity distribution was determined.
As in the single vortex case there are significant widths
associated with these most probable results.  The widths come
about because of the existance of other vortices as well as
fluctuations in the order parameter field.
There are, as shown in Fig. 2  , significant widths in the
probabilities due to the presence of other vortices and
fluctuations in the order parameter field.  For example,
as $x\rightarrow 0$, while the most probable relative velocity is
$4.38/R$, the half-width at half maximum for this quantity,
in these same units, is $2.08/R$.
In the large separation limit the probabilities become, as
expected, uncorrelated and each has the distribution of velocities
found previously\cite{MAZENKO97} for a single vortex.

The analysis here is built upon previous work on the ordering
kinetic of $O(n)$ symmetric systems.  The best available
theories\cite{2}
for the order parameter correlation function were built up in
the early 1990's and have led to the belief that we have a fairly
good understanding of how to calculate the associated scaling
function.   It also has become
clear that the order parameter correlation function or structure
factor is a rather structureless quantity which does not give a great
deal of direct information about the underlying disordering agents.
This led Liu and Mazenko\cite{LIU92b} to look directly at the correlations
between defects in the scaling regime.  The key new element in this
work, as discussed in some detail below, was the realization that
the positions of the vortices could be labelled by the zeros of
the order parameter field which could, in turn, be mapped onto the
zeros of an auxiliary field $\vec{m}(\vec{x},t)$.
They were able to show,
following work by Halperin\cite{4}, how one could write explicit
expressions for the signed and unsigned vortex densities in terms
of the auxiliary field $\vec{m}(\vec{r},t)$. This
then avoids the technically defeating step normally encountered which
requires one to indentify the vortex positions.

The signed vortex
density correlation function was determined analytically in
Ref(\onlinecite{LIU92b}) in
terms of the variance of the auxiliary field
under the assumption the auxiliary field is gaussian.
This calculation left the auxiliary field correlation function
$f(x)$ undetermined.  Liu and Mazenko assumed that one could use
$f(x)$ determined from a treatment of the order parameter dynamics
{\bf away~from~the~defect~cores}.

The charged or {\bf signed} vortex autocorrelation function does not
separate out all
of the desired information since it mixes the correlations between
like and unliked signed vortices.  It is not difficult to introduce
an {\bf unsigned} vortex autocorrelation function.  Between the
signed and unsigned autocorrelation functions one can construct
linear combinations which give the vortex-vortex and vortex-antivortex
correlation functions.  Unfortunately, for technical reasons
it is more difficult to determined the uncharged autocorrelation
function.  Only recently have these difficulties been overcome by
Mazenko and Wickham\cite{MWann}.  They found the results, expected on physical
grounds, that there is a depletion zone at short distances for the
vortex-vortex correlation function indicating repulsion.
Simulations\cite{MONDELLO90} and experiments\cite{NAGAYA95}
also show a depletion zone
at short distances for like-signed defects.  This is expected on
physical grounds since like-signed defects repel one another.
There is a clear discrepancy between theory and simulation results
at short-scaled distances.  The theory shows a monotonic
behavior as the separation distance goes to zero.  The simulation,
however, shows a maximum at short separation distances and then
falls rapidly to zero.  The depletion zone seen in this case in
the simulations is harder to understand physically since the pair
is attractive and headed toward annihilation.  While the theory
satisfies the sum rule implied by topological charge conservation, it
does not appear that this general constraint is satisfied by the
simulations.  It appears that the short distance behavior in the simulations
is contaminated by the choice of a vortex core distance which is comparable
to distances associated with the unphysical depletion zone.

It seems clear that it would be desirable to supplement this
information on the spatial correlation of vortices with
information concerning vortex velocities. It was
recently shown by the
author\cite{MAZENKO97} that
one could write down an explicit expression for the velocities
associated with point defects in terms of the order parameter field.
A key ingrediant in this development is the identification of a continuity
equation satisfied by the signed or charged vortex density.  This
continuity equation
gives a fundamental expression for conservation of topological charge in
the system. Using the gaussian closure assumption one can
determine the single vortex velocity distribution $P[\vec{v}_{1}]$.
The
most interesting physical result is that there is a large velocity
tail which was interpreted there as arising from the high velocities
in the late stages of vortex anti-vortex annihilation.
Bray\cite{Bray97} has used scaling arguments to obtain the same
large velocity tail.  The existance of these large velocities
will be supported by the calculation carried out here.

One common and concerning element in the calculations of
defect correlation functions and defect velocity distributions
is the
requirement that the auxiliary field scaled correlation function,
$f(x)$, be analytic as a function of $x$ for short scaled distances.
For example the fourth order gradient $(-\nabla^{4}_{x}f(x))|_{x=0}$
enters naturally into the analysis of $P[\vec{v}_{1}]$.  The need for
analyticity in x for $f(x)$ is not naturally consistent with the
simplest self-consistent analysis of $f(x)$ following a treatment
of the order parameter correlation function.
Mazenko and Wickham\cite{MWfluc} recently showed that one
can construct the theory so that $f(x)$ is analytic in $x$ for
small $x$, but this was at the expense of making the properties of the
order parameter correlation function worse\cite{OP}.

The tension between using the order parameter dynamics to determine
$f(x)$ and the requirement that $f(x)$ be analytic in order to
treat defect dynamics has been, to a degree, releaved by the very
recent work of Mazenko and Wickham\cite{MWOJK}.  They used the newly proposed
continuity equation for topological charge to derive the equation
satisfied by the auxiliary field correlation function under the
circumstances that the field is constrained to be near a defect
core.  As discussed briefly in Sect.III.c of this paper, they find
the clean result that the auxiliary field correlation function
determined in this manner satisfies a linear equation.  This
result is self-consistent with the assumption that the auxilary
field is gaussian.  The solution of the associated linear equation
has the Ohta, Jasnow and Kawasaki\cite{OJK} (OJK)  form
\be
f(x)=e^{-\frac{1}{2}x^{2}}
\label{eq:3}
\ee
which
is clearly analytic in the small x regime.
They argue in Ref.\onlinecite{MWOJK}
that the use of the
gaussian assumption in determining defect dyanamics, such as
$P(\vec{v}_{1})$, has a stronger
fundamental justification than in the case of the determination of the
order parameter correlation function.
In the calculation of the two-vortex velocity probability distribution
presented here it is assumed that the
order parameter field can be replaced
by a gaussian field in those portions of space near a vortex core
and the associted auxiliary field correlation function is of the
OJK form.

\section{Order Parameter Dynamics}

The system studied here has a defect dynamics generated by the
time-dependent Ginzburg-Landau (TDGL)
model satisfied by a nonconserved $n$-component vector order
parameter $\vec{\psi}(\vec{r},t)$:
\be
\frac{\partial \vec{\psi}}{\partial t}=\vec{K}\equiv
-\Gamma \frac{\delta F}{\delta \vec{\psi}}  +\vec{\eta}
\label{eq:2}
\ee
where $\Gamma $ is a kinetic coefficient,  $F$ is a Ginzburg-Landau
effective free energy assumed to be of the form
\be
F=\int ~d^{d}r \biggl( \frac{c}{2}(\nabla \vec{\psi})^{2}
+V(|\vec{\psi}|)\biggr)
\ee
where $c > 0$ and the potential $V$ is assumed to be of the
$O(n)$ symmetric
degenerate double-well
form. Since only these properties
of $V$ will be important in what follows we need not be more specific in our
choice for $V$ \cite{POTENTIAL}.
$\vec{\eta }$ is a thermal noise which is related to $\Gamma$
by a
fluctuation-dissipation theorem.
We assume that the quench is from a high temperature ($T_{I}>T_{c}$)
where the system is disordered to zero temperature where the noise
can be set to zero ($\vec{\eta}=0$).
It is believed that our final results are independent of the exact nature of
the initial state, provided it is a disordered state.

It is well
established that for late times following a quench from the disordered to the
ordered phase the dynamics obey scaling and the system can be described in
terms of a single growing length $L(t)$, which is characteristic of the
spacing between defects. In this scaling regime the order-parameter
correlation function has a universal equal-time scaling form

\be
\label{EQ:OPCOR}
C(12) \equiv \langle \OP (1) \cdot \OP (2) \rangle= \psi_{0}^{2} \F(x)
\ee
where $\psi_{0}$ is the magnitude $\psi = |\OP|$ of the order-parameter in the
ordered phase.  Here we use the short-hand notation where $1$ denotes
$({\bf r_{1}},t_{1})$. The scaled length $x$ is defined as
$\vec{x} = (\vec{r}_{1}-\vec{r}_{2})/L(t)$ where
$L(t) \sim t^{1/2}$
for the non-conserved models considered here.

In previous work on the order parameter scaling function it was
important to make a mapping of the order parameter $\vec{\psi}$
onto an auxiliary field $\vec{m}$ with the key requirement that
{\bf away} from defect cores
\be
\vec{\psi}=\psi_{0}\hat{m}  ~~~~~~ (A)
\ee
for the lowest-energy defects having unit topological charge.
Physically one expects that {\bf near} the defect cores
\be
\vec{\psi}=a\vec{m}+b(\vec{m})^{2}\vec{m}  +\cdot\cdot\cdot ~~~~~ (B)
\ee
for charge $\pm 1$ defects where $a$ and $b$ are constants.
In the theory for the order parameter
correlations it is property A which is important.  In the theory of
defect motion, as presented here, it is property B which is
important.  In this paper only property B enters into the analysis
since we always work near the defect cores.
To complete the definition of the model one must specify the form of
the probability distribution for the auxiliary field $\M$.
The simplest choice is a gaussian probability distribution for
$\M$ with
\be
\langle m_{\nu } (1) m_{\nu '} (2) \rangle = \delta_{\nu \nu '} \mbox{ }
C_{0} (12).
\ee
The system is assumed to be statistically isotropic and homogeneous so
$C_{0}(12)$ is invariant under interchange of its spatial indices.
In the scaling regime at equal times ($t_{1}¥=t_{2}¥=t$)
we introduce the auxiliary field autocorrelation function mentioned in
the introduction
\be
f(x)= C_{0}¥(\vec{r}_{1}¥t,\vec{r}_{2}¥t)/S_{0}(t)
\ee
and $S_{0}¥(t)= C_{0}(11)$ grows as $L^{2}¥(t)$ with
time after the quench.

\section{Topological Defects}

\subsection{Densities}

It has been emphasized in Refs.(\onlinecite{LIU92b})
that the signed or charged point
(n=d) defect density can be written in the form
\be
\rho ({\bf R},t)=\delta(\vec{ \psi}({\bf R},t)){\cal D}(\vec{R},t)
\label{eq:3.1}
\ee
where the Jacobian obtained with the change of variables
from the set of vortex positions to the zeros of the
field $\vec{\psi}$ is defined by:
\be
{\cal D}({\bf R},t)=\frac{1}{n!}\epsilon_{\mu_{1},\mu_{2},...,\mu_{n}}
\epsilon_{\nu_{1},\nu_{2},...,\nu_{n}}
\nabla_{\mu_{1}}\psi_{\nu_{1}}
\nabla_{\mu_{2}}\psi_{\nu_{2}}....
\nabla_{\mu_{n}}\psi_{\nu_{n}}
\label{eq:det}
\ee
where $\epsilon_{\mu_{1},\mu_{2},...,\mu_{n}}$ is the
$n$-dimensional fully anti-symmetric tensor and
summation over repeated indices is implied. The key point is that the
zeros of the order parameter $\vec{\psi}$ locates the positions of the
vortices.
The unsigned  density, $n({\bf R},t)$,
is given by
\be
n ({\bf R},t)=\delta(\vec{ \psi}({\bf R},t))|{\cal D}(\vec{R},t)|  ~~~.
\ee
The charged vortex correlation function is given by
\be
C_{\rho\rho}({\bf R},t)=<\rho({\bf R},t)\rho(0,t)>
{}~~~.
\ee
While the unsigned vortex correlation
function is given by
\be
C_{nn}({\bf R},t)= \langle  n({\bf R},t)n({\bf 0},t)\rangle ~~~.
\ee
It is shown in Ref.(\onlinecite{MWann}) that the vortex-vortex and
vortex-antivortex correlation functions can be expressed in terms
of $C_{\rho\rho}$ and $C_{nn}$.
$C_{\rho\rho}$ was evaluated in Ref.(\onlinecite{LIU92b})
using the gaussian closure
approximation.
As shown in Ref.(\onlinecite{MWann}),
the evaluation of $C_{nn}$ in this same approximation
is technically much more difficult
than the calculation of $C_{\rho\rho}$
because of the absolute value sign in the definition of the
unsigned defect density n.

\subsection{Conservation of Topological Charge}

It was shown in Ref.(\onlinecite{MAZENKO97}) that the charged
vortex density satisfies the continuity
equation,
\be
\frac{\partial\rho}{\partial t}=\nabla_{\beta}\Biggl[\delta
(\vec{\psi})J_{\beta}
^{(K)}\Biggr]
\label{eq:10}
\ee
where
\be
J_{\alpha}^{(K)}
=\frac{1}{(n-1)!}\epsilon_{\alpha,\mu_{2},...,\mu_{n}}
\epsilon_{\nu_{1},\nu_{2},...,\nu_{n}}
K_{\nu_{1}}
\nabla_{\mu_{2}}\psi_{\nu_{2}}....
\nabla_{\mu_{n}}\psi_{\nu_{n}}~~~.
\ee
A
key point here is that $J_{\beta}^{(K)}$ is multiplied by the vortex
locating
$\delta$-function.  This means that one can replace $\vec{K}$
in $\vec{J}^{(K)}$ by the
part of
$\vec{K}$ which does not vanish as $\vec{\psi}\rightarrow 0$.  Thus
in the case of a nonconserved order parameter one can replace
$J_{\beta}^{(K)}$ in the
continuity equation by
\be
J_{\beta}^{(2)}
=\frac{\Gamma c}{(n-1)!}\epsilon_{\beta,\mu_{2},...,\mu_{n}}
\epsilon_{\nu_{1},\nu_{2},...,\nu_{n}}
\nabla^{2}\psi_{\nu_{1}}
\nabla_{\mu_{2}}\psi_{\nu_{2}}....
\nabla_{\mu_{n}}\psi_{\nu_{n}}~~~.
\ee
Because of the standard form of the continuity equation Eq.(\ref{eq:10}),
it is clear that one
can identify the vortex velocity field as
\be
v_{\alpha}=-\frac{J_{\alpha}^{(2)}}{{\cal D}} ~~~.
\label{eq:18}
\ee
This form for
the velocity field is used inside expressions
multiplied by the vortex
locating  $\delta$-function.

\subsection{Use of Topological Charge Conservation to Determine the
Auxilliary Field Correlation Function}

In previous work\cite{MAZENKO90,LIU92a} a rather successful scheme
has been developed for
evaluating the order-parameter correlation function,
${\cal F}(x)$ and, in turn,
the auxilliary field correlation function $f(x)$.
As indicated in the introduction this leads to the problem that
the auxiliary field correlation function is rendered nonanalytic
as a function of $x$ for small $x$.  Mazenko and Wickham\cite{MWOJK} have
recently shown that this problem can be addressed in a different way.
Rather than using the order
parameter equation of motion to determine order-parameter correlation function
they used the continuity equation for topological charge
to determine the auxiliary field correlation function.
As in the rest of this paper we use the fact that in quantities
like $\rho$ and $\vec{v}$, we can replace
$\vec{\psi}\rightarrow\vec{m}$ everywhere.  Then we can
determine $f(x)$ by satisfying
\be
\frac{\partial}{\partial t}<\rho(1)\rho(2)>=
\nabla^{\beta}_{(1)}<\delta(\vec{\psi} (1))J_{\beta}^{(2)}(1)\rho(2)>
+\nabla^{\beta}_{(2)}<\rho(1)\delta(\vec{\psi} (2))J_{\beta}^{(2)}(2)> ~~~.
\label{eq:19}
\ee
under the assumption
that $\vec{m}$ is a gaussian field.
The calculation of the left-hand-side of Eq.(\ref{eq:19})
amounts to the evaluation of $C_{\rho\rho}$.  This calculation
was
carried out in Ref.(\onlinecite{LIU92b}) and is
straightforward since $C_{\rho\rho}$
factorizes into a product of gaussian averages which can
be evaluated using standard methods.  The calculation of the
average over $J_{\beta}^{(2)}$ can be organized in a similar
fashion.  In the scaling regime, after an impressive set of
cancellations, one finds the rather simple result that
\be
-\mu x f'=\nabla^{2}f+n\frac{S^{(2)}}{\sigma} f
\ee
where
\be
S^{(2)}=\frac{1}{n^{2}}<(\nabla\vec{m})^{2}>
\label{eq:27}
\ee
\be
S_{0}=\sigma L^{2} ~~~.
\ee
and we introduce the constant
\be
\mu=\frac{L\dot{L}}{2\Gamma c}~~~.
\ee
This equation for $f$
is linear and
has the simple solution of the
OJK form
\be
f=e^{-\frac{\mu}{2}x^{2}}
\ee
with the conditions
\be
n\frac{S^{(2)}}{\sigma} =\left(-\nabla^{2}f\right)|_{x=0}=n\mu ~~~.
\ee
For simplicity we set $\mu =1$ which amounts to choosing
$L(t)=2\sqrt{\Gamma ct}$ and results in the result for $f$
given by Eq.(\ref{eq:3}).

\subsection{Vortex Velocities}

As an important application of the result Eq.(\ref{eq:18}) for
the vortex velocity field $\vec{v}$ consider the
velocity
probability distribution function defined by
\be
n_{0} P(\vec{v}_{1})\equiv
\langle n\delta (\vec{v}_{1}-\vec{v})\rangle
\ee
where $\vec{v}_{1}$ is a reference velocity,
$\vec{v}$ is given by Eq.(\ref{eq:18}),
$n $ is the unsigned defect density,
and $n_{0}=\langle n\rangle$.
$P[\vec{v}_{1}]$ was found in Ref.(\onlinecite{MAZENKO97}) to
be given by
\be
P(\vec{v}_{1})=\frac{\Gamma (\frac{n}{2}+1)}{(\pi \bar{v}^{2})^{n/2}}
\frac{1}{\Biggl( 1+(\vec{v}_{1})^{2}/\bar{v}^{2}\Biggr)^{(n+2)/2}}
\label{eq:25}
\ee
where the the characteristic velocity $\bar{v}$ is defined by
\be
\bar{v}^{2}=(\Gamma c)^{2}\frac{\bar{S}_{4}}{S^{(2)}}
\label{eq:26}
\ee
where $S^{(2)}$ is given by Eq.(\ref{eq:27})
and
\be
\bar{S}_{4}=\frac{1}{n}<(\nabla^{2}\vec{m})^{2}>
-\frac{(nS^{(2)})^{2}}{S_{0}} ~~~.
\ee
Using the OJK form for $f(x)$ we obtain
$S^{(2)}=\sigma$,
$\bar{S}_{4}=\frac{d\sigma}{\Gamma ct}$
and
$\bar{v}^{2}=\frac{d\Gamma c}{t}$.

\section{Calculation of The Two Vortex Velocity Probability Distribution}

\subsection{General Development}

The main quantity of interest in this paper is the two velocity
correlation function defined by
\be
C_{nn}P[\vec{v}_{1},\vec{v}_{2}]=<n(1)
\delta (\vec{v}_{1}-\vec{v}(1))
n(2)
\delta (\vec{v}_{2}-\vec{v}(2))> ~~~.
\ee
where $\vec{v}_{1}$ and $\vec{v}_{2}$ are external labels while
the $\vec{v}(i)$, for $i=1,2$ , is expressed in terms of the order
parameter field $\psi (i)$ via Eq.(\ref{eq:18}).
$C_{nn}P[\vec{v}_{1},\vec{v}_{2}]$
is normalized such that the integrals over
$\vec{v}_{1}$ and $\vec{v}_{2}$ gives the unsigned defect density
correlation function $C_{nn}$  which was determined previously in
Ref.(\onlinecite{MWann}).

The first step in the
evaluation of $P[\vec{v}_{1},\vec{v}_{2}]$ is to notice that
we can
replace $\vec{\psi}$ by $\vec{m}$ in the
expressions for the unsigned vortex density and the velocity.
Next we need
to show that it
can be expressed in terms of an average
over a reduced probability distribution.  In the Appendix
we introduce the fields
\be
W_{i}[\xi ,\vec{b}]\equiv \delta (\vec{m}(i))
\delta (\vec{b}(i)-\nabla_{i}^{2}\vec{m}(i))
\prod_{\mu ,\nu =1}^{n}\delta (\xi_{\mu}^{\nu}(i)-\nabla_{\mu}^{(i)}m_{\nu}(i))
\ee
which have the normalization
\be
\int d^{n}b(i)\prod_{\mu ,\nu =1}^{n}d\xi_{\mu}^{\nu}(i)
W_{i}[\xi ,\vec{b}]=\delta (\vec{m}(i)) ~~~.
\ee
Using this result we can insert the factors of $W_{1}W_{2}$ into the
expression for $P[\vec{v}_{1},\vec{v}_{2}]$ and use the properties of
the $\delta$-function to replace all gradients and Laplacians
of $\vec{m}$ with the associated values constrained by the
multiplying $\delta$-function to obtain
\bea
\nonumber
C_{nn}P[\vec{v}_{1},\vec{v}_{2}]=\int \prod_{i=1}^{2}
\left[ d^{n}b(i)\prod_{\mu ,\nu =1}^{n}d\xi_{\mu}^{\nu}(i)
|{\cal D}(\xi (i))|
\delta (\vec{v}_{i}-\vec{v}(\xi (i),\vec{b}(i)))\right]
G_{2}(\xi ,\vec{b})
\eea
where
\be
G_{2}(\xi ,\vec{b})\equiv
<W_{1}[\xi ,\vec{b}]W_{2}[\xi ,\vec{b}]>
\ee
\be
{\cal D}(\xi )=\frac{1}{n!}\epsilon_{\mu_{1},\mu_{2},...,\mu_{n}}
\epsilon_{\nu_{1},\nu_{2},...,\nu_{n}}
\xi_{\mu_{1}}^{\nu_{1}}
\xi_{\mu_{2}}^{\nu_{2}}....
\xi_{\mu_{n}}^{\nu_{n}}
\ee
\be
v_{\alpha}(\xi (i),\vec{b}(i))=-\frac{J_{\alpha}^{(2)}(\xi (i),\vec{b}(i))}
{{\cal D}(\xi (i) )}
\ee
with
\be
J_{\alpha}^{(2)}(\xi (i),\vec{b}(i))=
\frac{\Gamma c}{(n-1)!}
\epsilon_{\alpha,\mu_{2},...,\mu_{n}}
\epsilon_{\nu_{1},\nu_{2},...,\nu_{n}}
b_{\nu_{1}}(i)
\xi_{\mu_{2}}^{\nu_{2}}(i)...
\xi_{\mu_{n}}^{\nu_{n}}(i)  ~~~.
\ee
The gaussian average giving $G_{2}(\xi ,\vec{b})$ is worked out explicitly in
the Appendix.  In the course of this calculation it is required that one
make a change of variables from $\xi_{\mu}^{\nu}(i)$ to a new set
$t_{\mu}^{\nu}(i)$  given by
\be
\xi_{\mu}^{\nu} (i) = \hat{R}^{\beta}_{\mu} t_{\beta}^{\nu} (i)
\ee
and  $\hat{R}^{\beta}_{\mu}$ is an othonormal matrix with the
additional
property that $\hat{R}^{(1)}_{\mu}=\hat{R}_{\mu}$ where
$\hat{R}_{\mu}$ is the unit vector pointing from vortex 2 to vortex
1.  Since $det (\hat{R}) =1$ the change of variables from $\xi$ to
$t$ is simple,
\be
\prod_{\nu\mu j}d\xi_{\mu}^{\nu}(j)=
\prod_{\nu\mu j}dt_{\mu}^{\nu}(j)
\ee
and
\be
{\cal D}(\xi (j)) ={\cal D}(t (j)) ~~~.
\ee
The one place in this change of variables where one must show care
is for the current
$\vec{J}^{(2)}$.  We have
\bea
\nonumber
J_{\alpha}^{(2)}(\xi (i),\vec{b}(i))=
\frac{\Gamma c}{(n-1)!}
\epsilon_{\alpha,\mu_{2},...,\mu_{n}}
\epsilon_{\nu_{1},\nu_{2},...,\nu_{n}}
b_{\nu_{1}}(i)
\hat{R}^{\beta_{2}}_{\mu_{2}}t_{\beta_{2}}^{\nu_{2}}(i)...
\hat{R}^{\beta_{n}}_{\mu_{n}}t_{\beta_{n}}^{\nu_{n}}(i) \\
=\frac{\Gamma c}{(n-1)!}\epsilon_{\alpha,\mu_{2},...,\mu_{n}}
\hat{R}^{\beta_{2}}_{\mu_{2}},...,\hat{R}^{\beta_{n}}_{\mu_{n}}
\epsilon_{\nu_{1},\nu_{2},...,\nu_{n}}b_{\nu_{1}}(i)
t_{\beta_{2}}^{\nu_{2}}(i),...,t_{\beta_{n}}^{\nu_{n}}(i) ~~~.
\eea
Clearly if we multiply this expression by $\hat{R}^{\beta_{1}}_{\alpha}$ and
sum over $\alpha$ we obtain
\bea
\nonumber
\hat{R}^{\beta_{1}}_{\alpha}J_{\alpha}^{(2)}(\xi (i),\vec{b}(i))
=\frac{\Gamma c}{(n-1)!}\hat{R}^{\beta_{1}}_{\alpha}
\epsilon_{\alpha,\mu_{2},...,\mu_{n}}
\hat{R}^{\beta_{2}}_{\mu_{2}},...,\hat{R}^{\beta_{n}}_{\mu_{n}}
\epsilon_{\nu_{1},\nu_{2},...,\nu_{n}}b_{\nu_{1}}(i)
t_{\beta_{2}}^{\nu_{2}}(i),...,t_{\beta_{n}}^{\nu_{n}}(i)
\eea
\bea
\nonumber
=\frac{\Gamma c}{(n-1)!}(det~\hat{R})~\epsilon_{\beta_{1},\beta_{2},...,\beta_{n}}
\epsilon_{\nu_{1},\nu_{2},...,\nu_{n}}b_{\nu_{1}}(i)
t_{\beta_{2}}^{\nu_{2}}(i),...,t_{\beta_{n}}^{\nu_{n}}(i)
\eea
\be
=J_{\alpha}^{(2)}(t (i),\vec{b}(i))~~~.
\ee
Multiplying by $\hat{R}^{\beta_{1}}_{\mu}$, summing over
$\beta_{1}$ and using
the orthonormality of the matrix $\hat{R}$ gives
\be
J_{\mu}^{(2)}(\xi (i),\vec{b}(i))
=\hat{R}^{\beta_{1}}_{\mu}J_{\beta_{1}}^{(2)}(t (i),\vec{b}(i))~~~.
\ee
Because of the rotational invariance of the $d$-dimensional
$\delta$-function we have
\be
\delta (\vec{v}_{i}-\vec{v}(\xi (i),\vec{b}(i)))=
\delta (\vec{u}(i)-\vec{v}(t (i),\vec{b}(i)))
\ee
where
\be
u(i)_{\mu}=\hat{R}_{\beta}^{\mu}v_{i,\beta}~~~.
\ee
Thus the $\mu =1$ component of $u_{\mu}$ is the longitudinal
component along
$\hat{R}$.  We then have that
\bea
C_{nn}P[\vec{v}_{1},\vec{v}_{2}]=\int\prod_{i=1}^{2}
\left[ d^{n}b (i)\prod_{\mu ,\nu =1}^{n}dt_{\mu}^{\nu}(i)
|{\cal D}(t (i))|
\delta (\vec{u}(i)-\vec{v}(t(i),\vec{b}(i)))\right]
G_{2}(t ,\vec{b})  ~~~.
\eea

The next step in the analysis is to perform the integration over
the $\vec{b}$ variables.  Toward this end we use the representation
\be
\delta (\vec{u}(i)-\vec{v}(t(i),\vec{b}(i)))
=\int \frac{d^{n}z(i)}{(2\pi)^{n}}
e^{-i\vec{u}(i)\cdot\vec{z}(i)}
e^{i\vec{v}(t(i),\vec{b}(i))\cdot\vec{z}(i)}
\ee
and we make explicit the $\vec{b}(i)$ dependence by writing
\be
\vec{v}(t(i),\vec{b}(i))\cdot\vec{z}(i)\equiv
a_{\nu}(i)b_{\nu}(i)
\ee
where
\bea
\nonumber
a_{\nu}(i)=-\frac{\Gamma c}{(n-1)!}\frac{1}{{\cal D}(t (i))}
z_{\alpha}(i)\epsilon_{\alpha\mu_{2}...\mu_{n}}
\epsilon_{\nu\nu_{2}...\nu_{n}}t_{\mu_{2}}^{\nu_{2}}(i)...
t_{\mu_{n}}^{\nu_{n}}(i)
\eea
\bea
=z_{\alpha}(i)N_{\nu\alpha}(i)
\eea
and
\be
N_{\nu\alpha}(i)=-\frac{\Gamma c}{(n-1)!}\frac{1}{{\cal D}(t (i))}
\epsilon_{\alpha\mu_{2}...\mu_{n}}
\epsilon_{\nu\nu_{2}...\nu_{n}}t_{\mu_{2}}^{\nu_{2}}(i)...
t_{\mu_{n}}^{\nu_{n}}(i)
\ee
Next one must make the $\vec{b}$ dependence of $G_{2}(t,\vec{b})$
explicit.
We have from the Appendix that
\be
G_{2}(\xi ,\vec{b})=G_{T}(t_{T})G_{L}(\vec{b},\vec{t}_{L}) ~~~.
\ee
The transverse part of $G_{2}$ does not depend on $\vec{b}(i)$,
while
the longitudinal contribution can be written
as
\be
G_{L}(\vec{b},\vec{t}_{L})
=\frac{1}{(2\pi )^{3n}}\frac{1}{(det ~ M)^{n/2}}
e^{-\frac{1}{2}\sum_{\alpha ,\beta =1}^{6}\vec{h}_{\alpha}
\cdot \vec{h}_{\beta} (M^{-1})_{\alpha \beta}} ~~~.
\ee
Where the matrix $M$ is discussed in detail in the Appendix
and the $\vec{h}_{\alpha}$ are defined by Eqs.(\ref{eq:A.48}
-\ref{eq:A.52}) .
Using the explicit expressions for the $\vec{h}_{\alpha}$
we can write
\be
\sum_{\alpha ,\beta =1}^{6}\vec{h}_{\alpha}
\cdot \vec{h}_{\beta} (M^{-1})_{\alpha \beta}
=S_{b}\sum_{i}\vec{b}(i)^{2}+2C_{b}\vec{b}(1)\cdot\vec{b}(2)
+2\sum_{i}\vec{b}(i)\cdot\vec{S}(i)+S_{L}^{0}
\sum_{i}\vec{t}_{L}^{2}(i)+2C_{L}^{0}\vec{t}_{L}(1)\cdot\vec{t}_{L}(2)
\ee
where we have defined
\be
S_{b}=(M^{-1})_{33}=(M^{-1})_{44}
\ee
\be
C_{b}=(M^{-1})_{34}=(M^{-1})_{43}
\ee
\be
\vec{S}(1)=(M^{-1})_{35}\vec{t}_{L}(1)+(M^{-1})_{36}\vec{t}_{L}(2)
\ee
\be
\vec{S}(2)=-(M^{-1})_{36}\vec{t}_{L}(1)-(M^{-1})_{35}\vec{t}_{L}(2)
\ee
\be
S_{L}^{0}=(M^{-1})_{55}=(M^{-1})_{66}
\ee
\be
C_{L}^{0}=(M^{-1})_{56}=(M^{-1})_{65}  ~~~.
\ee
The matrix inverse$M^{-1}$ is also discussed in detail in the Appendix.
It is convenient to define
\be
G_{L}^{0}(\vec{b},\vec{t}_{L})
=\frac{1}{(2\pi )^{3n}}\frac{1}{(det ~ M)^{n/2}}
e^{-\frac{1}{2}\left[S_{L}^{0}
\sum_{i}\vec{t}_{L}^{2}(i)+2C_{L}^{0}\vec{t}_{L}(1)\cdot\vec{t}_{L}(2)\right]}
\ee
and the $b$-dependence is explict when we write
\be
G_{2}(\xi ,\vec{b})=G_{T}(t_{T})G_{L}^{0}(\vec{t}_{L})
e^{-\frac{1}{2}\left[S_{b}\sum_{i}\vec{b}(i)^{2}
+2C_{b}\vec{b}(1)\cdot\vec{b}(2)
+2\sum_{i}\vec{b}(i)\cdot\vec{S}(i)\right]}~~~.
\ee
Putting this together we have
\be
C_{nn}P[\vec{v}_{1},\vec{v}_{2}]=\int\prod_{i=1}^{2}
\left[ \prod_{\mu ,\nu =1}^{n}dt_{\mu}^{\nu}(i)
\frac{d^{n}z(i)}{(2\pi)^{n}}|{\cal D}(t (i))|\right]
G_{T}(t_{T})G_{L}^{0}(\vec{t}_{L})
e^{-i\sum_{i}\vec{u}(i)\cdot\vec{z}(i)}
J_{0}
\ee
where the $\vec{b}(i)$ integrations are isolated in
\be
J_{0}=\int \prod_{i}d^{n}b(i)e^{i\sum_{i}\vec{b}(i)
\cdot\left[\vec{a}(i)+i\vec{S}(i)\right]}
e^{-\frac{1}{2}\left[S_{b}\sum_{i}\vec{b}(i)^{2}
+2C_{b}\vec{b}(1)\cdot\vec{b}(2)\right]}~~~.
\ee
This integration is of the standard gaussian form.  If we define
\be
\vec{A}(i)=\vec{a}(i)+i\vec{S}(i)
\ee
then we have the result
\be
J_{0}=(2\pi )^{n}\left(\frac{\gamma_{b}}{S_{b}}\right)^{n}
e^{-\frac{1}{2}\frac{\gamma_{b}^{2}}{S_{b}}{\cal Q}}
\ee
where
\be
{\cal Q}=
\sum_{i=1}^{2}\vec{A}^{2}(i)-2f_{b}\vec{A}(1)\cdot\vec{A}(2)~~~.
\ee
and
\be
\gamma_{b}=(1-f_{b}^{2})^{-1/2}
\ee
\be
f_{b}=C_{b}/S_{b} ~~~.
\ee
The next step is to do the $\vec{z}(i)$ integrations.  We can
highlight the $z$ dependence if we remember that
\be
a_{\nu}(i)
=z_{\alpha}(i)N_{\nu\alpha}(i) ~~~.
\ee
It is then a matter of straighforward algebra to show that
\be
{\cal Q}=-\sum_{i}\vec{S}^{2}(i)+2f_{b}\vec{S}(1)\cdot\vec{S}(2)
+\frac{S_{b}}{\gamma_{b}^{2}}
\sum_{\alpha\beta}\sum_{ij}z_{\alpha}(i)E_{\alpha\beta}(ij)
z_{\beta}(j)
\ee
where
\be
E_{\alpha\beta}(ij)=\frac{\gamma_{b}^{2}}{S_{b}}
\Omega_{\alpha\beta}(ij)[\delta_{ij}
-f_{b}\delta_{j,i+1}]
\label{eq:80}
\ee
and
\be
\Omega_{\alpha\beta}(ij)=\sum_{\nu}N_{\nu\alpha}(i)
N_{\nu\beta}(j) ~~~.
\ee
Here we have introduced the convenient notation that the index
$i$ is periodic, so that if $i=2$ then $i+1=1$.
Putting these results together we have
\bea
\nonumber
C_{nn}P[\vec{v}_{1},\vec{v}_{2}]=\int\prod_{i=1}^{2}
\left[ \prod_{\mu ,\nu =1}^{n}dt_{\mu}^{\nu}(i)
|{\cal D}(t (i))|\right]
G_{T}(t_{T})G_{L}^{0}(\vec{t}_{L})\\
\times (2\pi )^{n}\left(\frac{\gamma_{b}}{S_{b}}\right)^{n}
e^{\frac{1}{2}\frac{\gamma_{b}^{2}}{S_{b}}
\left[\sum_{i}\vec{S}^{2}(i)-2f_{b}\vec{S}(1)\cdot\vec{S}(2)\right]}
J_{1}
\eea
where the $z$ integration is given explicitly by
\be
J_{1}=\int\prod_{i=1}^{2}\left[\frac{d^{n}z(i)}{(2\pi)^{n}}\right]
e^{-i\sum_{i}\vec{U}(i)\cdot\vec{z}(i)}
e^{-\frac{1}{2}
\sum_{\alpha\beta}\sum_{ij}z_{\alpha}(i)E_{\alpha\beta}(ij)
z_{\beta}(j)}
\ee
where
\be
\vec{U}_{i}=\vec{u}(i)+\vec{d}(i)
\ee
and
\be
d_{\alpha}(i)=\frac{\gamma_{b}^{2}}{S_{b}}
\sum_{\nu =1}^{n}
N_{\nu\alpha}(i)\left[S_{\nu}(i)-f_{b}S_{\nu}(i+1)\right] ~~~.
\ee
$J_{1}$ is again of the standard form for a gaussian integral so
\be
J_{1}=\frac{1}{(2\pi )^{n/2}}\frac{1}{\sqrt{det~E}}
exp\left[-\frac{1}{2}U_{\alpha}(i)(E^{-1})_{\alpha\beta}(ij)
U_{\beta}(j)\right]
\ee
where, again, we need the determinant and the inverse of a matrix,
in this case $E$.
Let us look at the inverse first.  If we note the important result
(used in Ref.(\onlinecite{MAZENKO97}))
\be
N_{\sigma\alpha}(i)t_{\alpha}^{\nu}(i)=-\Gamma c\delta_{\sigma\nu}~~~,
\ee
where we do not sum on i, then
\be
\Omega_{\alpha\beta}(ij)t_{\beta}^{\sigma}(j)=-\Gamma cN_{\alpha\sigma}
\ee
where we do not sum on j.  These identities suggest that we try
a solution for $E^{-1}$ of the form
\be
(E^{-1})_{\alpha\beta}(ij)=\sum_{\nu}t_{\alpha}^{\nu}(i)
e_{ij}t_{\beta}^{\nu}(j)
\ee
with $e_{ij}$ to be determined. Inserting this ansatz into
the equation defining the inverse we easily find that
\be
e_{ij}=\frac{1}{(\Gamma c)^{2}}[S_{b}\delta_{ij}
+C_{b}\delta_{j,i+1}] ~~~.
\ee
In computing
$det~E$ we use the fact that
\be
det~E=\frac{1}{det~ (E^{-1})}
\ee
and that $E^{-1}$ can be written as the matrix product
\be
(E^{-1})_{\alpha\beta}(ij)=\sum_{\nu\nu' k\ell}t_{\alpha}^{\nu}(i)\delta_{ik}
e_{k\ell }\delta_{\nu\nu'}\delta_{\ell j}t_{\beta}^{\nu'}(j)
\ee
so that
\bea
\nonumber
det~ E^{-1}=det~t_{\alpha}^{\nu}(i)\delta_{ik}
{}~det~\left( e_{k\ell }\right)\delta_{\nu\nu'}
{}~det~\delta_{\ell j}t_{\beta}^{\nu'}(j) \\
={\cal D}(t(1)){\cal D}(t(2))(det ~ e )^{n}~{\cal D}(t(1)){\cal D}(t(2))
\eea
and
\be
det~e =\frac{1}{(\Gamma c)^{4}}(S_{b}^{2}-C_{b}^{2})
=\frac{1}{(\Gamma c)^{4}}\frac{S_{b}^{2}}{\gamma_{b}^{2}} ~~~.
\ee
Pulling all of this together leads to the result
\bea
\nonumber
C_{nn}P[\vec{v}_{1},\vec{v}_{2}]=\frac{1}{(\Gamma c)^{2n}}
\int\left[\prod_{i=1}^{2}
\prod_{\mu ,\nu =1}^{n}dt_{\mu}^{\nu}(i)
{\cal D}^{2}(t (i))\right]
G_{T}(t_{T})G_{L}^{0}(\vec{t}_{L}) \\
\times
e^{\frac{1}{2}\frac{\gamma_{b}^{2}}{S_{b}}
\left[\sum_{i}\vec{S}^{2}(i)-2f_{b}\vec{S}(1)\cdot\vec{S}(2)\right]}
exp\left[-\frac{1}{2}\sum_{ij\alpha\beta}
U_{\alpha}(i)(E^{-1})_{\alpha\beta}(ij)
U_{\beta}(j)\right]
\label{eq:103}
\eea
and the important point is that one does not have an absolute
value  sign left in the Jacobian factors.
Turn next to the argument of the exponential in the last line of
Eq.(\ref{eq:103}).
After a substantial amount of algebra we find
\bea
\nonumber
-\frac{\gamma_{b}^{2}}{S_{b}}
\left[\sum_{i}\vec{S}^{2}(i)-2f_{b}\vec{S}(1)\cdot\vec{S}(2)\right]
+\sum_{ij\alpha\beta}U_{\alpha}(i)(E^{-1})_{\alpha\beta}(ij)
U_{\beta}(j)\\
=\sum_{ij}\vec{{\cal V}}(i)\cdot\vec{{\cal V}}(j)e_{ij}
-\frac{2}{\Gamma c}\sum_{i}\vec{{\cal V}}(i)\cdot\vec{S}(i)
\eea
where
\be
{\cal V}_{\nu}(i)=\sum_{\alpha}u_{\alpha}(i)t_{\alpha}^{\nu}(i)~~~.
\ee
Putting this together leads to
\be
C_{nn}P[\vec{v}_{1},\vec{v}_{2}]=\frac{1}{(2\pi )^{3n}}
\frac{1}{(\Gamma c)^{2n}}\frac{1}{(det~M)^{n/2}}
\left(\frac{\gamma_{T}}{2\pi S_{T}}\right)^{n(n-1)}J_{2}
\ee
where the final integration is over the matrices $t_{\alpha}^{\beta}(i)$:
\be
J_{2}=
\int\left[\prod_{i=1}^{2}
\prod_{\mu ,\nu =1}^{n}dt_{\mu}^{\nu}(i)
{\cal D}^{2}(t (i))\right]e^{-\frac{1}{2}A(t)}
\ee
where
\bea
\nonumber
A(t)=
\frac{\gamma_{T}^{2} }{ S_{T}}  \sum_{\mu = 2}^{n}
\sum_{\nu = 1}^{n}\left[
\sum_{i=1}^{2}[t_{\mu}^{\nu} (i) ] ^{2} - 2 f_{T}
t_{\mu}^{\nu} (1) t_{\mu}^{\nu} (2)\right]
+S_{L}^{0}
\sum_{i}\vec{t}_{L}^{2}(i)\\
+2C_{L}^{0}\vec{t}_{L}(1)\cdot\vec{t}_{L}(2)
+\sum_{ij}\vec{{\cal V}}(i)\cdot\vec{{\cal V}}(j)e_{ij}
-\frac{2}{\Gamma c}\sum_{i}\vec{{\cal V}}(i)\cdot\vec{S}(i) ~~~.
\eea
$A(t)$ is clearly a quadratic form in the matrix $t_{\mu}^{\nu}(i)$.
The quantities $S_{T}$, $f_{T}$ and $\gamma_{T}$ govern the
transverse modes and are defined by Eqs.(\ref{eq:A57}
-\ref{eq:A60}) in the Appendix.
After sufficient rearrangement $A(t)$ can be written in the final form
\be
A(t)=\sum_{ij\alpha\beta\nu}t_{\alpha}^{\nu}(i)
W_{\alpha\beta}(ij)t_{\beta}^{\nu}(j)
\ee
where the matrix $W$ plays a central role in the theory and is
given by the manifestly symmetric form
\be
W_{\alpha\beta}(ij)=\delta_{\alpha\beta}d_{ij}^{\alpha}
+u_{\alpha}(i)\Omega_{ij}^{(1)}u_{\beta}(j)
+u_{\alpha}(i)\Omega_{ij}^{(2)}L_{\beta}(j)
+L_{\alpha}(i)\Omega_{ij}^{(2)}u_{\beta}(j)
\ee
where
\be
d_{ij}^{\alpha}=\delta_{ij}d_{\alpha}+\delta_{i+1,j}d_{\alpha}^{c}
\ee
\be
d_{\alpha}=\delta_{\alpha ,L}S_{L}^{0}+(1-\delta_{\alpha ,L})
\frac{\gamma_{T}^{2}}{S_{T}}
\ee
\be
d_{\alpha}^{c}=\delta_{\alpha ,L}C_{L}^{0}+(1-\delta_{\alpha ,L})
f_{T}\frac{\gamma_{T}^{2}}{S_{T}}
\ee
\be
\Omega_{ij}^{(1)}=\delta_{ij}\frac{S_{b}}{(\Gamma c)^{2}}
+\delta_{i+1,j}\frac{C_{b}}{(\Gamma c)^{2}}
\ee
\be
\Omega_{ij}^{(2)}=-\delta_{ij}\frac{(M^{-1})_{35}}{\Gamma c}
+\delta_{i+1,j}\frac{(M^{-1})_{36}}{\Gamma c}
\ee
and
\be
L_{\alpha}(i)=\delta_{\alpha ,L}\eta_{i}
\ee
where
\be
\eta_{1}=-\eta_{2}=1 ~~~.
\ee
The final integration over the matrices $t_{\alpha}^{\nu}(i)$
is not of the standard form evaluated so far but instead there is
the polynomial ${\cal D}^{2}(1){\cal D}^{2}(2)$ multiplying the
gaussian in the integrand.  It is technically important that there are no
absolute value signs left in this expression and the integral
can be evaluated by introducing a field $g_{\alpha}^{\nu}(i)$ which
couples to $t_{\alpha}^{\nu}(i)$ via
\be
-\frac{1}{2}A(t)\rightarrow -\frac{1}{2}A(t)
+\sum_{\alpha\nu i}g_{\alpha}^{\nu}(i)t_{\alpha}^{\nu}(i) ~~~.
\ee
If we consider
\be
J_{2}(g)=
\int\left[\prod_{i=1}^{2}
\prod_{\mu ,\nu =1}^{n}dt_{\mu}^{\nu}(i)
{\cal D}^{2}(t (i))\right]e^{-\frac{1}{2}A(t)}
e^{\sum_{\alpha\nu i}g_{\alpha}^{\nu}(i)t_{\alpha}^{\nu}(i)} ~~~,
\ee
then any polynomial can be generated by taking derivatives with
respect to $g$.
Using the explicit expressions for the ${\cal D}^{2}(t (i))$
we have
\bea
\nonumber
J_{2}(g)=
\sum_{\mu_{1}...\mu_{n}=1}^{n}\epsilon_{\mu_{1}...\mu_{n}}
\frac{\partial}{\partial g_{\mu_{1}}^{1}(1)}...
\frac{\partial}{\partial g_{\mu_{n}}^{n}(1)}
\sum_{\mu_{1}'...\mu_{n}'=1}^{n}\epsilon_{\mu_{1}'...\mu_{n}'}
\frac{\partial}{\partial g_{\mu_{1}'}^{1}(1)}...
\frac{\partial}{\partial g_{\mu_{n}'}^{n}(1)} \\
\times
\sum_{\nu_{1}...\nu_{n}=1}^{n}\epsilon_{\nu_{1}...\nu_{n}}
\frac{\partial}{\partial g_{\nu_{1}}^{1}(2)}...
\frac{\partial}{\partial g_{\nu_{n}}^{n}(2)}
\sum_{\nu_{1}'...\nu_{n}'=1}^{n}\epsilon_{\nu_{1}'...\nu_{n}'}
\frac{\partial}{\partial g_{\nu_{1}'}^{1}(2)}...
\frac{\partial}{\partial g_{\nu_{n}'}^{n}(2)}
J_{3}(g)
\label{eq:142}
\eea
where
\be
J_{3}(g)=
\int\left[\prod_{i=1}^{2}
\prod_{\mu ,\nu =1}^{n}dt_{\mu}^{\nu}(i)\right]
e^{-\frac{1}{2}A(t)}
e^{\sum_{\alpha\nu i}g_{\alpha}^{\nu}(i)t_{\alpha}^{\nu}(i)} ~~~.
\ee
$J_{3}(g)$ is now of the standard form and we have
\be
J_{3}(g)=\frac{(2\pi)^{n^{2}}}{(det~W)^{n/2}}
exp\left[\frac{1}{2}\sum_{ij\alpha\beta\nu}
g_{\alpha}^{\nu}(i)\Lambda_{\alpha\beta}(ij)
g_{\beta}^{\nu}(j)\right]
\ee
where $\Lambda_{\alpha\beta}(ij)$ is the matrix inverse of $W$.  It is
then straightforward to take the derivatives with respect to $g$ and then
set $g$ to zero to obtain
\be
J_{2}=
\frac{(2\pi)^{n^{2}}}{(det~W)^{n/2}}J_{4}
\ee
where
\bea
\nonumber
J_{4}=
\sum_{\mu_{1}...\mu_{n}=1}^{n}\epsilon_{\mu_{1}...\mu_{n}}
\sum_{\mu_{1}'...\mu_{n}'=1}^{n}\epsilon_{\mu_{1}'...\mu_{n}'}
\sum_{\nu_{1}...\nu_{n}=1}^{n}\epsilon_{\nu_{1}...\nu_{n}}
\sum_{\nu_{1}'...\nu_{n}'=1}^{n}\epsilon_{\nu_{1}'...\nu_{n}'}\\
\times \prod_{\sigma =1}^{n}\left[
\Lambda_{\mu_{\sigma}\mu_{\sigma}'}(11)
\Lambda_{\nu_{\sigma}'\nu_{\sigma}}(22)
+\Lambda_{\mu_{\sigma}\nu_{\sigma}'}(12)
\Lambda_{\mu_{\sigma}'\nu_{\sigma}}(12)
+\Lambda_{\mu_{\sigma}\nu_{\sigma}}(12)
\Lambda_{\mu_{\sigma}'\nu_{\sigma}'}(12)\right]
\eea
and the final result is
\be
C_{nn}P[\vec{v}_{1},\vec{v}_{2}]=\frac{1}{(2\pi )^{3n}}
\frac{1}{(\Gamma c)^{2n}}\frac{1}{(det~M)^{n/2}}
\left(\frac{\gamma_{T}}{2\pi S_{T}}\right)^{n(n-1)}
\frac{(2\pi)^{n^{2}}}{(det~W)^{n/2}}J_{4}~~~.
\label{eq:130}
\ee
Finally all of the integrals have been evaluated.  What is left is to
evaluate the determinant and matrix inverse of $W$.

\subsection{Large x Limit}

As a check on the preceeding analysis it is useful to work out the large scaled
distance limit where we expect the probability distribution to factorize
into the product for each of the tagged vortices.  In this limit,
using the results from
Table I,
we find that the matrices entering $W$ are in a diagonal form
\be
d_{ij}^{\alpha}=D\delta_{ij}
\ee
where
\be
D=\frac{1}{\sigma }
\ee
\be
\Omega_{ij}^{(1)}=\frac{L^{2}}{2d\sigma (\Gamma
c)^{2}}\delta_{ij}
\ee
and\be
\Omega_{ij}^{(2)}=0 ~~~.
\ee
The matrix $W$ can then be written in the partially diagonal form
\be
W_{\alpha\beta}(ij)=\delta_{ij}\left[\delta_{\alpha\beta}D
+\bar{u}_{\alpha}(i)\bar{u}_{\beta}(i)\right]
\ee
where
\be
\bar{u}_{\alpha}(i)=\frac{Lu_{\alpha}(i)}{\Gamma
c\sqrt{2n\sigma}}~~~.
\ee
Notice that there is no longer a difference between the longitudinal
and transverse directions as expected.
The inverse matrix $\Lambda$ then satsifies the equation
\be
D\Lambda_{\alpha\beta}(ij)+\bar{u}_{\alpha}(i)
\sum_{\mu}\bar{u}_{\mu}(i)\Lambda_{\mu\beta}(ij)
=\delta_{\alpha\beta}\delta_{ij}~~~.
\ee
This equation is in the form of a trap which can first be solved to obtain
\be
\sum_{\mu}\bar{u}_{\mu}(i)\Lambda_{\mu\beta}(ij)=
\delta_{ij}\frac{\bar{u}_{\beta}(i)/D}{1+\sum_{\mu}\bar{u}_{\mu}^{2}(i)/D}
\ee
and the full inverse is given by
\be
\Lambda_{\alpha\beta}(ij)=\delta_{ij}\left[D^{-1}\delta_{\alpha\beta}
-\frac{\bar{u}_{\alpha}(i)\bar{u}_{\beta}(i)}
{D^{2}\left[1+\sum_{\mu}\bar{u}_{\mu}^{2}(i)/D\right]}\right]~~~.
\ee

With these results it is easy to see that the quantity $J_{4}$ can be
written in the simple form
\be
J_{4}=(n!)^{2}det ~\Lambda(11))det~ \Lambda(22)
\ee
\be
det~W=det ~W(11))det~ W(22)
\ee
and
\be
det ~\Lambda(ii))=\frac{1}{det ~W(ii))} ~~~.
\ee

It is then easy to see that
\be
det~W[ii]=D^{n}[1+v_{i}^{2}/\bar{v}^{2}]
\ee
where $\bar{v}^{2}$ is given by Eq.(\ref{eq:26}) .  If we then carefully
keep track of all the factors we see that Eq.(\ref{eq:130}) reduces to
\be
\lim_{x\rightarrow\infty}C_{nn}P[\vec{v}_{1},\vec{v}_{2}]
= n_{0}P[\vec{v}_{1}]
n_{0}P[\vec{v}_{2}]
\ee
where $n_{0}P[\vec{v}_{i}]$ is given by Eq.(30) in
Ref.(\onlinecite{MAZENKO97}) and, after proper normalization,
leads to the
expression for the single vortex velocity probability distribution
given by Eq.(\ref{eq:25}).

\subsection{n=d=2 Case}

The general expression for $C_{nn}P[\vec{v}_{1},\vec{v}_{2}]$, is
complicated.
Let
us restrict ourselves here to the case of $n=d=2$ where $det~W$ and
$J_{4}$
can be evaluated explicitly.
Let us define
\be
det~ A(ij)=A_{11}(ij)A_{22}(ij)-A_{12}(ij)A_{21}(ij)
\ee
where the matrix $A_{\alpha\beta }(ij)$ is either $W$ or its inverse
$\Lambda$.  We can also define
\bea
\nonumber
Q_{A}=\left[A_{11}(22)A_{21}(21) -A_{21}(22)A_{11}(21)\right]
\left[A_{12}(11)A_{22}(12) -A_{22}(11)A_{12}(12)\right] \\
\nonumber
-\left[A_{12}(22)A_{21}(21) -A_{22}(22)A_{11}(21)\right]
\left[A_{12}(11)A_{21}(12) -A_{22}(11)A_{11}(12)\right]\\
\nonumber
-\left[A_{11}(22)A_{22}(21) -A_{21}(22)A_{12}(21)\right]
\left[A_{11}(11)A_{22}(12) -A_{21}(11)A_{12}(12)\right] \\
+\left[A_{12}(22)A_{22}(21) -A_{22}(22)A_{12}(21)\right]
\left[A_{11}(11)A_{21}(12) -A_{21}(11)A_{11}(12)\right]  ~~~~.
\eea
In terms of these quantities we have
\be
det~W =det~W(11)det~W(22) +det~W(12)det~W(21) +Q_{W}
\ee
and
\be
J_{4}=4\left[det~\Lambda (11)det~\Lambda (22)
+3det~\Lambda (12)det~\Lambda (21) +Q_{\Lambda}\right] ~~~.
\ee
It is clear that the last nontrivial step before evaluating
$C_{nn}P[\vec{v}_{1},\vec{v}_{2}]$ is to determine
$\Lambda_{\alpha\beta }(ij)$.  This will be carried out in general in
section IV.E   ,  however most of the important physics can be extracted
in the problem by considering the simple case where the transverse
velocities are both zero.  In this case one can make substantial
analytical  progress.

\subsection{Zero Transverse Velocities}

Before tackling the complete determination of $P[\vec{v}_{1},\vec{v}_{1}]$
it is very instructive to study the much simpler
case where the transverse velocities are set to zero.  This case is of
interest not only because it is simple but also because it is the most
probable situation.  The most likely situation is that each of
the two tagged
vortices will have zero transverse velocity.

If the transverse velocities are zero then the problem simplifies since
the matrix $W$ reduces to the diagonal form
\be
W_{\alpha\beta}(ij)=\delta_{\alpha\beta}D_{ij}^{\alpha}
\ee
where
\be
D_{ij}^{\alpha}=d_{ij}^{\alpha}+\delta_{\alpha ,L}
\left[
u(i)\Omega_{ij}^{(1)}u(j)
+u(i)\Omega_{ij}^{(2)}\eta_{j}
+\eta_{i}\Omega_{ij}^{(2)}u(j)\right]
\ee
and $u(i)=u_{L}(i)$.  Clearly the longitudinal and transverse degrees of
freedom are uncoupled and we have after some manipulations
\be
D_{ij}^{T} =d^{T}\left[\delta_{ij}+f_{T}\delta_{j,i+1}\right]
\ee
while
\be
D_{ij}^{L} =a_{i}\delta_{ij}+b\delta_{j,i+1}
\ee
where
\be
a_{i}=(M^{-1})_{55}+(M^{-1})_{33}\bar{u}(i)^{2}-2\eta_{i}
(M^{-1})_{35}\bar{u}(i) ~~~,
\label{eq:138}
\ee
\be
b=(M^{-1})_{56}+(M^{-1})_{34}\bar{u}(1)\bar{u}(2)
+(M^{-1})_{36}\left[\bar{u}(2)-\bar{u}(1)\right]
\label{eq:139}
\ee
and
\be
\bar{u}(i)=\frac{u(i)}{\Gamma c} ~~~.
\ee
In this case we see that $W$ is a relatively simple matrix.  We need
its determinant and then its inverse on the way to evaluating the
quantity $J_{4}$.  The matrix of $W$ for general $n$ is simply given by
\be
det~W=(det~D^{T})^{n-1}(det~D^{L})
\ee
where
\be
det~D^{T} = (d^{T})^{2}(1-f_{T}^{2})
=\frac{1}{S_{T}^{2}-C_{T}^{2}}
\ee
and
\be
det~D^{L} =a_{1}a_{2}-b^{2} ~~~.
\ee
The quantity $det~D^{L}$ is key in the development and we shall
return to it soon.  First we need to evaluate the inverse of $W$
to complete the calculation.  In this case this involves the solution
of the equation
\be
\sum_{k}D_{ik}^{\alpha}\Lambda_{\alpha\beta}(kj)=\delta_{\alpha\beta}
\delta_{ij}
\ee
which is easily found to be given by
\be
\Lambda_{\alpha\beta}(ij)=\delta_{\alpha\beta}
\left[\delta_{\alpha ,L}\Lambda_{L}(ij)
+\delta_{\alpha ,T}\Lambda_{T}(ij)\right]
\ee
where
\be
\Lambda_{L}(ij)=\frac{1}{det~D^{L}}\left[
a_{i+1}\delta_{ij}-b\delta_{j,i+1}\right]
\ee
and
\be
\Lambda_{T}(ij)=\frac{1}{det~D^{T}}d^{T}\left[
\delta_{ij}-f_{T}\delta_{j,i+1}\right] ~~~.
\ee
Using these results one can work out the quantity $J_{4}$ for the
case $n=d=2$ with the result
\be
J_{4}=4\left[ det~\Lambda (11)det~\Lambda (22)+
3\left(det ~\Lambda (22)\right)^{2}
-\Lambda_{L}(22)\Lambda_{L}(11)\Lambda_{T}^{2}(12)
-\Lambda_{T}(22)\Lambda_{T}(11)\Lambda_{L}^{2}(12)\right]
\ee
where we have used the fact that the matrices
$\Lambda_{L}(ij)$ and $\Lambda_{T}(ij)$
are symmetric.  Putting in
the explicit forms for $\Lambda$ we obtain
\be
J_{4}=\frac{4}{(det~D^{T})^{2}(det~D^{L})^{2}}
\left[det~D^{L}det~D^{T}+
2b^{2}(d^{T}f_{T})^{2}\right] ~~~.
\ee
Putting all of this together for $n=d=2$ and the transverse
velocities zero we have
\be
C_{nn}P[\vec{v}_{1},\vec{v}_{2}]=\frac{4}{(\Gamma c)^{6}}
\frac{1}{det~M}\left[\frac{1}{(det~D^{L})^{2}det~D^{T}}
+\frac{2b^{2}C_{T}^{2}}{(det~D^{L})^{3}}\right] ~~~.
\ee

Clearly the next step is the explicit evaluation of
$det~D^{L}$.  Using the  expressions for $a_{i}$ and
$b$ given by Eqs.(\ref{eq:138}) and (\ref{eq:139}) we obtain,
after some algebra, that
\bea
\nonumber
det~D^{L}=\frac{\sigma^{4}}{D_{O}D_{E}}\Bigg[
\bar{\gamma}_{0}
+\frac{\bar{\gamma}_{1}}{2}\left(\tilde{u}(1)-\tilde{u}(2)\right)
+\bar{\gamma}_{A}\left(\tilde{u}^{2}(1)+\tilde{u}^{2}(2)\right)\\
+\bar{\gamma}_{B}\tilde{u}(1)\tilde{u}(2)
+\frac{\bar{\gamma}_{3}}{2}
\tilde{u}(1)\tilde{u}(2)\left(\tilde{u}(1)-\tilde{u}(2)\right)
+\bar{\gamma}_{4}\tilde{u}^{2}(1)\tilde{u}^{2}(2)\Bigg]
\eea
where the scaled longitudinal velocities are defined by
\be
\tilde{u}(i)=\bar{u}(i)L ~~~.
\ee
and
\be
\bar{\gamma}_{0}=-\kappa_{O}^{(2)}\kappa_{E}^{(2)}
\ee
\be
\bar{\gamma}_{1}=2\left[(1-f)\kappa_{O}^{(1)}\kappa_{E}^{(2)}
+(1+f)\kappa_{E}^{(1)}\kappa_{O}^{(2)}\right]
\ee
\be
\bar{\gamma}_{A}=(1-f)\frac{D_{E}}{8\sigma^{3}}
+(1+f)\frac{D_{O}}{2\sigma^{3}}+\frac{\bar{\gamma}_{2}}{4}
\ee
\be
\bar{\gamma}_{B}=(1-f)\frac{D_{E}}{4\sigma^{3}}
+(1+f)\frac{D_{O}}{\sigma^{3}}-\frac{\bar{\gamma}_{2}}{2}
\ee
\be
\bar{\gamma}_{2}=(1-f)\kappa_{O}^{(0)}\kappa_{E}^{(2)}
-(1+f)\kappa_{E}^{(0)}\kappa_{O}^{(2)}
-4(1-f^{2})\kappa_{O}^{(1)}\kappa_{E}^{(1)}
\ee
\be
\bar{\gamma}_{3}=2(1-f^{2})\left[\kappa_{O}^{(1)}\kappa_{E}^{(0)}
-\kappa_{E}^{(1)}\kappa_{O}^{(0)}\right]
\ee
\be
\bar{\gamma}_{4}=(1-f^{2})\kappa_{E}^{(0)}\kappa_{O}^{(0)}  ~~~.
\ee
where the $\kappa$'s  are given as functions of $f$ by
Eqs.(\ref{eq:A92}-\ref{eq:A97}) in appendix A. $D_{O}$ and $D_{E}$
are given as functions of $f$ by
Eqs.(\ref{eq:A84}-\ref{eq:A85}) in appendix A.
Note the relationship
$2\bar{\gamma}_{A}-\bar{\gamma}_{B}=\bar{\gamma}_{2}$.
Notice the crucial result that after rescaling the velocities by a
factor of L, the time dependence drops out of $det~D^{L}$.  This will
eventually lead to the result that scaling holds for the probability
distribution at late times if we rescale velocities in this manner.

We also need to express the quantity $b$ in terms of the
$\kappa$'s.  It is convenient to write
\be
b=b_{O}+b_{E}
\ee
where
\be
b_{O}=\frac{\sigma^{2}}{4D_{O}}\left[-\kappa_{O}^{(2)}
-(1-f)\kappa_{O}^{(0)}\tilde{u}(1)\tilde{u}(2)
+(1-f)\kappa_{O}^{(1)}\left(\tilde{u}(2)-\tilde{u}(1)\right)\right]
\ee
\be
b_{E}=\frac{\sigma^{2}}{D_{E}}\left[-\kappa_{E}^{(2)}
+(1+f)\kappa_{E}^{(0)}\tilde{u}(1)\tilde{u}(2)
+(1+f)\kappa_{E}^{(1)}\left(\tilde{u}(2)-\tilde{u}(1)\right)\right] ~~~.
\ee
The last ingrediant needed to evaluate the probability distribution is
\be
C_{T}=-\frac{C_{0}'}{R}=-\sigma\frac{f'(x)}{x}=\sigma f(x)
\ee
using the OJK form for $f(x)$  in the last step.

Let us look first at
the small x limit.  Since the
OJK form for $f(x)$ is easily expanded in a power series in $x$
and we can extract to leading order in $x$:
$\bar{\gamma}_{0}=32x^{4}$,
$\bar{\gamma}_{1}=-24x^{5}$,
$\bar{\gamma}_{2}=\frac{26}{3}x^{6}$,
$\bar{\gamma}_{3}=-\frac{4}{3}x^{7}$,
$\bar{\gamma}_{4}=\frac{x^{8}}{12}$,
$\bar{\gamma}_{A}=\frac{3}{8}x^{2}$,
$\bar{\gamma}_{B}=2\bar{\gamma}_{A}$.
We also need
\be
D_{E}=16x^{2}\sigma^{3}
\ee
\be
D_{O}=\sigma^{3}\frac{x^{8}}{48} ~~~.
\ee
Notice that the $\bar{\gamma}_{A}$ and $\bar{\gamma}_{B}$ dominate
the expression for $det~D^{L}$ in the small x limit
and we can write to leading orders in
$x$
\be
det~D^{L}=\frac{\sigma^{4}}{D_{O}D_{E}}
\left[\bar{\gamma}_{0}+\bar{\gamma}_{A}\left(\tilde{u}(1)
+\tilde{u}(2)\right)^{2}\right]
\ee
plus terms which are higher order in $x$.  We can write this in the
more convenient form
\be
det~D^{L}=\frac{\sigma^{4}\bar{\gamma}_{0}}{D_{O}D_{E}}
\left[1+V^{2}/V_{0}^{2}
\right]
\ee
where after some algebra we obtain
\be
\frac{\sigma^{4}\bar{\gamma}_{0}}{D_{O}D_{E}}=\frac{96}{\sigma^{6}x^{6}}
\ee
\be
V=\tilde{u}(1)
+\tilde{u}(2)
\ee
and
\be
V_{0}^{2}=\frac{\bar{\gamma}_{0}}{\bar{\gamma}_{A}}
=8x^{2} ~~~.
\ee
Similarly we find that for small $x$ that $b$ is dominated by
$b_{0}$ and given to leading order in $x$ by
$b=-\frac{384}{x^{4}\sigma }$.
We also need
$C_{T}=\sigma$
to leading order in $x$.
The probability distribution is dominated in the small $x$ limit by the
term proportional to $b^{2}$.  The other term is down by a factor of
$x^{4}$.  Putting all of this together we obtain
\be
C_{nn}P[\vec{v}_{1},\vec{v}_{2}]=
\left(\frac{\sigma^{2}}{\Gamma c }\right)^{6}
\frac{1}{(1+V^{2}/V_{0}^{2})^{3}} ~~~.
\ee
Then as $x\rightarrow 0$ we find, with increasing probability, that
\be
V=\tilde{u}(1)
+\tilde{u}(2)=0 ~~~.
\ee
This is just the physical statement that there is very low probability that
there is a nonzero momentum of the center of mass of the two tagged
vortices.  Thus all of the action is {\bf in} the center of mass
where we can set
\be
\tilde{u}(2)=-\tilde{u}(1)\equiv\tilde{u} ~~~.
\ee

If we return to the probability distribution for the case where the
COM momentum is zero, then
\be
det~D^{L}=\frac{\sigma^{4}}{D_{O}D_{E}}\left[
\bar{\gamma}_{0}
+\bar{\gamma}_{1}\tilde{u}
+\bar{\gamma}_{2}\tilde{u}^{2}
+\bar{\gamma}_{3}
\tilde{u}^{3}
+\bar{\gamma}_{4}\tilde{u}^{4}\right]
\ee
and
\be
b_{O}=\frac{\sigma^{2}}{4D_{O}}\left[-\kappa_{O}^{(2)}
+(1-f)\kappa_{O}^{(0)}\tilde{u}^{2}
+2(1-f)\kappa_{O}^{(1)}\tilde{u}\right]
\ee
\be
b_{E}=\frac{\sigma^{2}}{D_{E}}\left[-\kappa_{E}^{(2)}
-(1+f)\kappa_{E}^{(0)}\tilde{u}^{2}
+2(1+f)\kappa_{E}^{(1)}\tilde{u}\right] ~~~.
\ee

In the small $x$ limit these reduce to
\be
b=b_{O}+b_{E}=-\frac{384}{x^{4}\sigma }f_{\beta}
\ee
where
\be
f_{\beta}=
1-\frac{u}{4}+\frac{u^{2}}{48}
\ee
\be
det(D^{L})=\frac{96}{\sigma^{6}x^{6}}f_{L}
\ee
with
\be
f_{L}=1-\frac{3}{4}u+\frac{13}{48}u^{2}-\frac{1}{24}u^{3}+\frac{1}{384}u^{4}
\ee
and the scaled velocity is given by
\be
u=\tilde{u}x ~~~.
\ee
The probability distribution is given in the $x\rightarrow 0$ limit by:
\be
C_{nn}P[\vec{v}_{1},\vec{v}_{2}]=
\left(\frac{\sigma^{2}}{\Gamma c }\right)^{6}
\frac{f_{\beta}^{2}}{f_{L}^{3}} ~~~.
\ee
A key conclusion we can draw at this point is that it is only the
combination $u=\tilde{u}x$ which enters the probability distribution
with high probability as $x\rightarrow 0$.  Thus the relative
velocity increases as $1/x$ as $x\rightarrow 0$.
We plot
$f_{\beta}^{2}/f_{L}^{3}$ as a function of $u$ in Fig.2.
We that the most probable values of the relative velocity as
a function of x for small $x$ are given by:
\be
v_{L}\equiv \frac{\Gamma c}{L} \frac{\kappa}{x}
=\frac{\Gamma c\kappa}{R}
\ee
with $\kappa =2.19$ a pure number.  This is the result quoted in
the introduction.

If we then plot the two vortex velocity probability distribution
function for zero transverse velocities in the COM for general
$x$ as shown in Figs. 3 and 4,
we obtain the most probable relative velocities as a function
of x as shown in Fig.1.
The interpretation that this is the interaction between
vortices and anti-vortices holds only out to modest values of x
where the population of same signed vortices begins to appear
(See Ref.(\onlinecite{MWann})).

\subsection{General Evaluation}

The complete determination of the two-vortex velocity probability distribution
as a general function of $\vec{v}_{1}$ and $\vec{v}_{2}$ can be
carried out in the $n=d=2$ case if one can invert the matrix
$W_{\alpha\beta}(ij)$ to obtain its inverse $\Lambda_{\alpha\beta}(ij)$
defined by
\be
\sum_{\mu ,k}W_{\alpha\mu}(ik)\Lambda_{\mu\beta}(kj)=
\delta_{\alpha\beta}\delta_{ij} ~~~.
\ee
This inversion is a quite unpleasant task if one heads in the wrong
direction.
It is useful in order to make the development more transparent to
introduce a mixed operator notation where $W_{\alpha\beta}$ is an
operator in the space associated with the indices i and j
\be
W_{\alpha\beta}(ij)=<i|W_{\alpha\beta}|j> ~~~.
\ee
Then the matrix $W_{\alpha\beta}$ is given by
\be
W_{\alpha\beta}=\delta_{\alpha\beta}d^{\alpha}
+u_{\alpha}\Omega^{(1)}u_{\beta}
+u_{\alpha}\Omega^{(2)}L_{\beta}
+L_{\alpha}\Omega^{(2)}u_{\beta}
\ee
where $u_{\alpha}$ and $L_{\alpha}$ are diagonal operators
\be
<i|u_{\alpha}|j>=u_{\alpha}(i)\delta_{ij}
\ee
\be
<i|L_{\alpha}|j>=L_{\alpha}(i)\delta_{ij}  ~~~.
\ee
The key idea is that if we can write $W_{\alpha\beta}$ in the
form
\be
W_{\alpha\beta}=\delta_{\alpha\beta}D_{\alpha}
+P_{\alpha}\tilde{P}_{\beta} ~~~,
\label{eq:220}
\ee
where $\tilde{P}_{\beta}$ is the transpose of $P_{\alpha}$, then we
can carry out the inversion straight away.  Let us first show this and
then return to show that $W$ can be written in the assumed form.

We want to invert the equation
\be
\sum_{\mu }W_{\alpha\mu}\Lambda_{\mu\beta}=
\delta_{\alpha\beta} ~~~.
\ee
Inserting the assumed form, Eq.(\ref{eq:220}), for $W$ we obtain
\be
D^{\alpha}\Lambda_{\alpha\beta}
+P_{\alpha}\sum_{\mu}\tilde{P}_{\mu}\Lambda_{\mu\beta}=
\delta_{\alpha\beta} ~~~.
\ee
Multiplying from the left by the matrix inverse of $D_{\alpha}$ this
becomes
\be
\Lambda_{\alpha\beta}
+D_{\alpha}^{-1}P_{\alpha}\sum_{\mu}\tilde{P}_{\mu}\Lambda_{\mu\beta}=
D_{\alpha}^{-1}\delta_{\alpha\beta} ~~~.
\label{eq:223}
\ee
This equation is then in the form of a trap for the quantity
$\sum_{\mu}\tilde{P}_{\mu}\Lambda_{\mu\beta}$  .  Multiplying
Eq.(\ref{eq:223})
by $\tilde{P}_{\alpha}$ and summing over $\alpha$ we obtain a closed
equation for $\sum_{\mu}\tilde{P}_{\mu}\Lambda_{\mu\beta}$ which
has the solution
\be
\sum_{\mu}\tilde{P}_{\mu}\Lambda_{\mu\beta}
=[1+Q]^{-1}\sum_{\mu}\tilde{P}_{\beta}D_{\beta}^{-1}
\ee
where the $2\times 2$ symmetric matrix $Q$ is defined by
\be
Q=\sum_{\mu}\tilde{P}_{\mu}D_{\mu}^{-1}P_{\mu} ~~~.
\ee
This leads directly to the final result
\be
\Lambda_{\alpha\beta}=D_{\alpha}^{-1}\delta_{\alpha\beta}
-D_{\alpha}^{-1}P_{\alpha}[1+Q]^{-1}\tilde{P}_{\beta}D_{\beta}^{-1}
\ee
which is clearly symmetric.
This gives a practical expression for the inverse
once one has identified the matrices $D$ and $P$.

The key observation which allows one to write $W$ in the desired form
given by Eq.(\ref{eq:220}) is that the matrix
$\Omega^{(1)}_{ij}$ can be factorized
in the form
\be
\Omega^{(1)}_{ij}=\sum_{k}\omega_{ik}\tilde{\omega}_{kj}
\ee
where
\be
\omega_{ij}=\omega_{0}\delta_{ij}+\omega_{1}\delta_{j,i+1}
\ee
and
\be
\omega_{0}=\frac{\sqrt{S_{b}}}{2\Gamma c}\left[
\sqrt{1+f_{b}}+\sqrt{1-f_{b}}\right]
\ee
\be
\omega_{1}=\frac{\sqrt{S_{b}}}{2\Gamma c}\left[
\sqrt{1+f_{b}}-\sqrt{1-f_{b}}\right] ~~~.
\ee
Using this factorization result it is then easy to show that $W$ can
be written in the form Eq.(\ref{eq:220})  with
\be
D_{ij}^{\alpha}=d_{ij}^{\alpha}-\delta_{\alpha ,L}
\sum_{k}C_{L}(ik)\tilde{C}_{L}(kj) ~~~,
\ee
\be
P_{\alpha}(ij)=u_{\alpha}(i)\omega_{ij} +C_{\alpha}(ij)
\ee
with
\be
C_{\alpha}(ij) =L_{\alpha}(i)\sum_{k}\Omega_{ik}^{(2)}
\omega_{kj}^{-1} ~~~.
\ee
Combining these results one has an explicit expression for the two-vortex
velocity
probability distribution for arbitrary velocities.  The major
qualitative feature of including the transverse velocities is to allow
one to look at the widths of the distributions in the transverse
directions since we find the most probable configurations are
those where the transverse velocites of both vortices are zero.
These widths turn out to be comparable to those associated with the
longitudinal modes.

\section{Discussion}

In the analysis here we have looked at the correlation between vortices
regardless of their signs.  At short relative distances, where it
is unlikely to
have two vortices of the same sign, one can interpret the results in terms
of vortex-antivortex dynamics.  It is clear that one can go further,
as discussed by Mazenko and Wickham\cite{MWann}, and separate the probability
distribution into that  for vortex-vortex and vortex-antivortex
pairs.
The key idea , which is essentially
equivalent to that used in the case of spatial
correlations, that a factor of
${\cal P}_{+}(1)\equiv \frac{1}{2}(1+sgn ~{\cal D}(1))$
restricts one the positive
charge vortex sector, while
${\cal P}_{-}(1)=\frac{1}{2}(1-sgn ~{\cal D}(1))$ restricts
one to the negative charge anti-vortex sector.  Thus the probability
for vortex-vortex correlations is
\be
C_{vv}P_{vv}(12)=
<n(1)\delta (\vec{v}_{1}-\vec{v}(1)){\cal P}_{+}(1)
n(2)\delta (\vec{v}_{2}-\vec{v}(2)){\cal P}_{+}(2)\rangle~~~~.
\ee
The vortex-antivortex contribution is given by
\be
C_{av}P_{av}(12)=
<n(1)\delta (\vec{v}_{1}-\vec{v}(1)){\cal P}_{-}(1)
n(2)\delta (\vec{v}_{2}-\vec{v}(2)){\cal P}_{+}(2)\rangle~~~~.
\ee
These quantities can be multiplied out and, and after using
symmetry to show that the correlation between the signed and
unsigned quantities are zero, can be expressed in terms of
the probability distribution determined in this paper and
\be
C_{\rho\rho}P_{\rho\rho}(12)=
<\rho(1)\delta (\vec{v}_{1}-\vec{v}(1))
\rho(2)\delta (\vec{v}_{2}-\vec{v}(2))\rangle
\ee
which has not yet been computed.  It is expected that $P_{\rho\rho}(12)$
will be difficult to determine because of the addition factors of the
$sgn ~{\cal D}$.  The analysis will be essentially identicle in
structure up to eq.(125) with the expression for $J_{2}$
 showing the replacement
\be
{\cal D}^{2}(t(i))\rightarrow {\cal D}(t(i))|{\cal D}(t(i))| ~~~.
\ee
The resulting integral for $J_{2}$ can not then be represented
in the product form given by Eq.(\ref{eq:142}).  This remains a problem to
be solved.

In principle Eq.(\ref{eq:130}) givens an expression which
can be integrated over
all velocities to give $C_{u}$ and determine the overall normalization.
It is not clear how to do this analytically since the velocities
appear in a complicated fashion in $det~W$ and $J_{4}$.  A numerical
determination is quite feasible.

In this paper we have shown how one can make progress in an analysis
of the dynamics of point vortices in the context of phase ordering
kinetics.  The results include the effects of other vortices and
order parameter fluctuations on the dynamics of the tagged
vortices.  The results appear completely physical and the
determination of the relative velocity as short distances appears to
be a useful result.  The method used here appears to generalize easily
to the case of string defects.  This will be the subject of subsequent
work.

\appendix
\section*{Gaussian Average}

In this appendix we work out the gaussian average
\be
G_{2}(\xi ,\vec{b})=<W_{1}[\xi ,\vec{b}]W_{2}[\xi ,\vec{b}]> ~~~.
\ee
where the $W$'s are defined by
\be
W_{i}[\xi ,\vec{b}]\equiv \delta (\vec{m}(i))
\delta (\vec{b}(i)-\nabla_{i}^{2}\vec{m}(i))
\prod_{\mu ,\nu =1}^{n}\delta (\xi_{\mu}^{\nu}(i)-\nabla_{\mu}^{(i)}m_{\nu}(i))
\ee
where we have already assumed that $n=d$ in the product.

The first step in the evaluation of $G_{2}$ is to use the
Fourier representation for the
$\delta$-function to obtain
\be
W_{i}[\xi ,\vec{b}]=\int d\tilde{\Omega}[i]
e^{i\vec{q}_{i}\cdot\vec{m}(i)}
e^{-i\vec{s}_{i}\cdot \bigl(\vec{b}(i)-\nabla_{i}^{2}
\vec{m}(i)\bigr)}
e^{-i\sum_{\mu\nu}k_{\mu}^{\nu}(i)\bigl(\xi_{\mu}^{\nu}(i)
-\nabla_{\mu}m_{\nu}(i)\bigr)}
\ee
where we have defined
\be
d\tilde{\Omega}[i]=\frac{d^{n}q_{i}}{(2\pi)^{n}}
\frac{d^{n}s_{i}}{(2\pi)^{n}}
\prod_{\mu,\nu =1}^{n}\left[ \frac{dk_{\mu}^{\nu}(i)}{2\pi}\right]~~~.
\ee
We can rewrite this in the more useful form
\be
W_{i}[\xi ,\vec{b}]=\int d\tilde{\Omega}[i]
e^{-i\left[\vec{s}_{i}\cdot \vec{b}(i)
+\sum_{\mu\nu =1}^{n}k_{\mu}^{\nu}(i)\xi_{\mu}^{\nu}(i)\right]}
e^{\sum_{\alpha =1}^{n}\int d\bar{1}~H_{i}^{\alpha}(\bar{1})
m_{\alpha}(\bar{1})}
\ee
where
\be
\int ~d\bar{1} = \int d^{d}\bar{r}_{1}d\bar{t}_{1}
\ee
\be
H_{i}^{\alpha}(\bar{1})=i\left[ q_{i}^{\alpha}+s_{i}^{\alpha}
\nabla_{(i)}^{2}+\sum_{\mu =1}^{n}k_{\mu}^{\alpha}(i)
\nabla_{(i)}^{\mu}\right]\delta (\bar{1}i)
\ee
and $\nabla_{(i)}^{\mu}$ is the $\mu^{th}$ component of the
gradient acting on $\vec{r}_{i}$.
The average of interest can then be written as
\be
G_{2}(\xi ,\vec{b})=\int ~d\tilde{\Omega}[1]d\tilde{\Omega}[2]
e^{-i\sum_{i=1}^{2}\left[\vec{s}_{i}\cdot \vec{b}(i)
+\sum_{\mu\nu =1}^{n}k_{\mu}^{\nu}(i)\xi_{\mu}^{\nu}(i)\right]}
<exp\left[\sum_{i=1}^{2}\sum_{\alpha =1}^{n}\int d\bar{1}
{}~H_{i}^{\alpha}(\bar{1})
m_{\alpha}(\bar{1})\right]> ~~~.
\ee
The average is of the standard form for a gaussian average with
the result
\be
<exp\left[\sum_{i=1}^{2}\sum_{\alpha =1}^{n}\int d\bar{1}
{}~H_{i}^{\alpha}(\bar{1})
m_{\alpha}(\bar{1})\right]>=
e^{-\frac{1}{2}A_{0}}
\ee
where
\be
A_{0}=-\sum_{i,j=1}^{2}
\sum_{\alpha ,\beta =1}^{n}\int d\bar{1}\int d\bar{2}
H_{i}^{\alpha}(\bar{1})H_{j}^{\beta}(\bar{2}) C_{0}(\bar{1},\bar{2})
\delta_{\alpha ,\beta}
\ee
and we have used
\be
<m_{\alpha}(\bar{1})m_{\beta}(\bar{2})>=C_{0}(\bar{1},\bar{2})
\delta_{\alpha ,\beta} ~~~.
\ee

Inserting the expression for $H$ into $A_{0}$ we need the following definitions:
\be
C_{0}(ii)\equiv S_{0}
\ee
\be
\left[\nabla_{(i)}^{2}C_{0}(ij)\right]|_{i=j}\equiv -nS^{(2)}
\ee
\be
\left[\nabla_{(j)}^{2}\nabla_{(i)}^{2}C_{0}(ij)\right]|_{i=j}\equiv S^{(4)}
\ee
\be
\left[\nabla_{(j)}^{\nu}\nabla_{(i)}^{\mu}C_{0}(ij)\right]|_{i=j}
=\delta_{\mu\nu}S^{(2)}~~~.
\ee
Using the fact that $C_{0}(12)$ depends only on the magnitude of
$\vec{R}=\vec{r}_{1}- \vec{r}_{2}$
we convert all derivatives to those
with respect to $\vec{R}$:
\be
\nabla_{(1)}^{\mu}C_{0}(12)=C_{0}'\hat{R}_{\mu}
\ee
\be
\nabla_{(2)}^{\mu}C_{0}(12)=-C_{0}'\hat{R}_{\mu}
\ee
where the prime indicates a derivative with respect to $R$.
Going further, for $i=1$ and $2$,
\be
\nabla_{(i)}^{2}C_{0}(12)=\nabla_{R}^{2}C_{0}(R)=C_{0}''
+\frac{(d-1)}{R}C_{0}'
\ee
\be
\nabla_{(1)}^{\mu}\nabla_{(2)}^{\nu}C_{0}(12)
=-C_{0}''\hat{R}_{\mu}\hat{R}_{\nu}-\frac{C_{0}'}{R}
(\delta_{\mu\nu}-\hat{R}_{\mu}\hat{R}_{\nu})
\ee
\be
\nabla_{(2)}^{2}\nabla_{(2)}^{\mu}C_{0}(12)
=-p\hat{R}_{\mu}
\ee
\be
\nabla_{(1)}^{2}\nabla_{(1)}^{\mu}C_{0}(12)
=p\hat{R}_{\mu}
\ee
where
\be
p=C_{0}'''+\frac{(d-1)}{R}(C_{0}''-\frac{C_{0}'}{R})
=(\nabla_{R}^{2}C_{0}(R))' ~~~.
\ee

We see that it is then natural to use the coordinate system
parallel and orthogonal to $\vec{R}$.  Indeed we can introduce
the orthonormal set $\hat{R}_{\beta}^{\alpha}$
where
\be
\sum_{\alpha =1}^{n}\hat{R}_{\alpha}^{\mu}\hat{R}_{\alpha}^{\nu}
=\delta_{\mu\nu}
\ee
\be
\sum_{\mu =1}^{n}\hat{R}_{\alpha}^{\mu}\hat{R}_{\beta}^{\mu}
=\delta_{\alpha\beta}~~~.
\ee
The only other
thing we need to know about this set is that
\be
\hat{R}_{\beta}^{1} =\hat{R}_{\beta} ~~~.
\ee
Next we define
\be
W_{\beta}^{\alpha}(i)=\sum_{\mu =1}^{n}\hat{R}_{\mu}^{\beta}
k_{\mu}^{\alpha}(i)
\label{eq:A27}
\ee
which can be inverted to give
\be
k_{\mu}^{\alpha}(i)=\sum_{\beta =1}^{n}\hat{R}_{\mu}^{\beta}
W_{\beta}^{\alpha}(i)  ~~~.
\ee
In terms of this new set of variables
\be
\sum_{i,\mu ,\nu}\left(k_{\mu}^{\nu}(i)\right)^{2}
=\sum_{i,\mu ,\nu}\left(\sum_{\beta}\hat{R}_{\mu}^{\beta}
W_{\beta}^{\nu}(i)\right)\left(\sum_{\sigma}
\hat{R}_{\mu}^{\sigma}W_{\sigma}^{\nu}(i)\right)
=\sum_{i,\beta ,\nu}\left(W_{\beta}^{\nu}(i)\right)^{2} ~~~.
\ee
We then have
\be
A_{0}=A(\vec{q})+A(\vec{s})+A(\vec{W}_{L})+A_{T}(W)
+A_{c}(\vec{q},\vec{s},\vec{W}_{L})
\ee
where
\be
A(\vec{q})=S_{0}\sum_{i=1}^{2}\vec{q}_{i}~^{2}
+2C_{0}\vec{q}_{1}\cdot\vec{q}_{2}
\ee
\be
A(\vec{s})=S^{(4)}\sum_{i=1}^{2}\vec{s}_{i}~^{2}
+2(\nabla^{4}C_{0})\vec{s}_{1}\cdot\vec{s}_{2}
\ee
\be
A_{L}(\vec{W})=S^{(2)}\sum_{i=1}^{2}\vec{W}_{L}(i)^{2}
-2(C_{0}'')\vec{W}_{L}(1)\cdot\vec{W}_{L}(2)
\ee
where the longitudinal part of the tensor $W$ is defined by
\be
W_{L}^{\alpha}(i)=\sum_{\mu =1}^{n}\hat{R}_{\mu}^{1}k_{\mu}^{\alpha}(i)~~~.
\ee
The {\it transverse} contribution to $A_{0}$ is given by
\be
A_{T}(W)=\sum_{\mu =2}^{n}\sum_{\nu=1}^{n}
\left[S^{(2)}\sum_{i=1}^{2}(W_{\mu}^{\nu}(i))^{2}
-2(C_{0}'/R)W_{\mu}^{\nu}(1)W_{\mu}^{\nu}(2)\right]~~~.
\ee
The term {\it coupling} the set $\vec{q},\vec{s},\vec{W}_{L}$
is given by
\bea
\nonumber
A_{c}(\vec{q},\vec{s},\vec{W}_{L})=-2nS^{(2)}\sum_{i=1}^{2}
\vec{q}_{i}\cdot\vec{s}_{i}
+2(\vec{q}_{1}\cdot\vec{s}_{2}+\vec{q}_{2}\cdot\vec{s}_{1})
\nabla^{2}C_{0}\\
-2(p\vec{s}_{1}+C_{0}'\vec{q}_{1})\cdot\vec{W}_{L}(2)
+2(p\vec{s}_{2}+C_{0}'\vec{q}_{2})\cdot\vec{W}_{L}(1)
\eea
Notice that the {\it transverse} modes decouple from the longitudinal
set coupled in $A_{c}$.

It will be very useful for us to rewrite $A_{0}$ as a sum of a
transverse part, already written down, and a logitudinal part
which is a quadratic form in the vector
\be
\vec{\phi}_{\alpha}=
\left[\vec{q}(1),\vec{q}(2),\vec{s}(1),\vec{s}(2),
\vec{W}_{L}(1),\vec{W}_{L}(2)\right]
\ee
where we assume that the subscript $\alpha$ runs from $1$ to $6$.  We have
then
\be
A_{L}=\sum_{\alpha\beta}M_{\alpha\beta}
\vec{\phi}_{\alpha}\cdot\vec{\phi}_{\beta}
\ee
where the matrix $M$ is given explicitly by.
\be
M=\left (
\begin{array}{cccccc}
 S_{0}   & C_{0} & u_{1} & u_{2} &  0       & u_{4}\\
 C_{0}   & S_{0} & u_{2} & u_{1} & -u_{4}   & 0\\
 u_{1}   & u_{2} & S^{(4)} & C_{4} &  0       & u_{3}\\
 u_{2}   & u_{1} & C_{4} & S^{(4)}& -u_{3}  & 0\\
 0     & -u_{4}& 0     & -u_{3} & S^{(2)} & C_{2}\\
 u_{4} & 0     & u_{3} &  0    & C_{2}    & S^{(2)}
\end{array}
\right ) ~~~.
\ee
Various quantities entering the matrix $M$ are defined by
\be
C_{4}=\nabla^{4}C_{0}=\frac{\sigma}{L^{2}}\nabla^{4}_{x}f(x)
\ee
\be
S^{(4)}=\nabla^{4}C_{0}|_{R=0}=8\frac{\sigma}{L^{2}}
\ee
\be
C_{2}=-C_{0}''=-\sigma f''
\ee
\be
u_{1}=-nS^{(2)}=-n\sigma
\ee
\be
u_{2}=\nabla^{2}C_{0}=\sigma\nabla^{2}_{x}f(x)
\ee
\be
u_{3}=-p=-\frac{\sigma}{L}(\nabla^{2}_{x}f(x))'
\ee
\be
u_{4}=-C_{0}'=-L\sigma f'(x) ~~~.
\ee

The complete change of variable from $k$ to $W$ in Eq.(\ref{eq:A27})
requires noting that the
Jacobian taking one from $k$ to $W$ is one.
The argument of the exponential outside the average in Eq.(\ref{eq:A27})
can also be written
in terms of the set
$\vec{\phi}_{\alpha}$ and the transverse part of $W$.
Then one has
\be
\sum_{i=1}^{2}\left[\sum_{\mu ,\nu =1}^{n}
k_{\mu}^{\nu}(i)\xi_{\mu}^{\nu}(i)+\vec{s}_{i}\cdot\vec{b}(i)\right]
=\sum_{\alpha =1}^{6}\vec{h}_{\alpha}\cdot\vec{\phi}_{\alpha}
+\sum_{i=1}^{2}\sum_{\nu =1}^{n}\sum_{\mu =2}^{n}
W_{\mu}^{\nu}(i)t_{\mu}^{\nu}(i)
\ee
where
\be
\vec{h}_{1}=\vec{h}_{2}=0
\label{eq:A.48}
\ee
\be
\vec{h}_{3}=\vec{b}(1)
\label{eq:A.49}
\ee
\be
\vec{h}_{4}=\vec{b}(2)
\label{eq:A.50}
\ee
\be
\vec{h}_{5}=\vec{t}_{L}(1)
\label{eq:A.51}
\ee
\be
\vec{h}_{6}=\vec{t}_{L}(2)
\label{eq:A.52}
\ee
and
\be
t_{\mu}^{\nu}(i)\equiv \sum_{\beta =1}^{n}
R_{\mu}^{\beta}(i)\xi_{\beta}^{\nu}(i) ~~~.
\ee
We then have the result that $G_{2}$ factorizes into longitudinal and
transverse components:
\be
G_{2}(\xi ,\vec{b})=G_{T}(t_{T})G_{L}(\vec{b},\vec{t}_{L}) ~~~.
\label{eq:A.54}
\ee
First consider  the transverse contribution given by
\be
G_{T} (t_{T})  =   \int \left[\prod_{i=1}^{2}
\prod_{\nu =1}^{n} \prod_{\mu = 2}^{n}
\frac{d W_{\mu}^{\nu}(i)}{2 \pi}\right] \exp \Biggl[-\frac{1}{2} A_{T}(W_{T})
  - i \sum_{i=1}^{2}
\sum_{\nu = 1}^{n}
\sum_{\mu = 2}^{n} t_{\mu}^{\nu} (i) W_{\mu}^{\nu} (i) \Biggr] ~~~.
\ee
This is a standard Gaussian integral which can be evaluated
with the results
\be
G_{T}(t_{T})  =  \left( \frac{\gamma_{T}}{2 \pi S_{T}} \right)^{n(n-1)}
\exp - \frac{\gamma_{T}^{2} }{2 S_{T}} \left[ \sum_{\mu = 2}^{n}
\sum_{\nu = 1}^{n}\left[
\sum_{i=1}^{2}[t_{\mu}^{\nu} (i) ] ^{2} - 2 f_{T}
t_{\mu}^{\nu} (1) t_{\mu}^{\nu} (2)\right] \right]
\label{eq:A.56}
\ee
where
\be
S_{T}  =  S^{(2)}
\label{eq:A57}
\ee
\be
C_{T}  =  - \frac{C_{0}'}{R}
\label{eq:A58}
\ee
\be
f_{T}  =  \frac{C_{T}}{S_{T}}
\label{eq:A59}
\ee
and
\be
\gamma_{T}^{2}  =  (1 - f_{T}^{2}) ^{-1} .
\label{eq:A60}
\ee

The longitudinal contribution to $G_{2}$ is also a standard
gaussian integral:
\bea
\nonumber
G_{L}(\vec{b},\vec{t}_{L})=\int \prod_{i=1}^{2}\left[
\prod_{\alpha=1}^{6}\frac{d^{n}\phi_{\alpha}(i)}{(2\pi )^{n}}\right]
e^{\left[-i\sum_{\alpha =1}^{6}\vec{h}_{\alpha}\cdot\vec{\phi}_{\alpha}\right]}
e^{-\frac{1}{2}\sum_{\alpha\beta =1}^{6}M_{\alpha\beta}
\vec{\phi}_{\alpha}\cdot\vec{\phi}_{\beta}}\\
=\frac{1}{(2\pi )^{3n}}\frac{1}{(det ~ M)^{n/2}}
e^{-\frac{1}{2}\sum_{\alpha ,\beta =1}^{6}\vec{h}_{\alpha}
\cdot \vec{h}_{\beta} (M^{-1})_{\alpha \beta}}
\eea
Thus the determination of $G_{2}$ reduces to an evaluation of the
inverse and determinant of the matrix $M$ given above.
If we multiply $M$ from left and right by the matrix
\be
Q=\left (
\begin{array}{cccccc}
  1   & \frac{1}{2} & 0            & 0   &  0          & 0 \\
 -1   & \frac{1}{2} & 0            & 0   &  0          & 0 \\
  0   & 0           & \frac{1}{2}  & 1   &  0          & 0 \\
  0   & 0           & -\frac{1}{2} & 1   &  0          & 0 \\
  0   & 0           & 0            & 0   & \frac{1}{2} & -1\\
  0   & 0           & 0            & 0   & \frac{1}{2} & 1
\end{array}
\right )
\ee
then the new matrix
\be
\tilde{M}_{\alpha \beta}=\sum_{\mu\nu}Q_{\mu\alpha}
Q_{\nu\beta}M_{\mu\nu}
\ee
has the block diagonal form
\be
\tilde{M}=\left (
\begin{array}{cccccc}
 2(S_{0}-C_{0})&0& u_{1}-u_{2}&0 &  u_{4}       & 0\\
0&\frac{1}{2}(S_{0}+C_{0})& 0 & u_{1}+u_{2} & 0 & u_{4}\\
u_{1}-u_{2}&0&\frac{1}{2}(S^{(4)}-C_{4})&0 &\frac{1}{2}u_{3}&0\\
0&u_{1}+u_{2}&0 &2(S^{(4)}+C_{4})&0 & 2u_{3}\\
 u_{4} & 0& \frac{1}{2}u_{3}&0&\frac{1}{2}(S^{(2)}+C_{2})&0\\
 0 & u_{4}&0&2u_{3}& 0& 2(S^{(2)}-C_{2})
\end{array}
\right ) ~~~.
\ee
It is easy to see that the inverse for $M$ can be expressed in
terms of the inverse of $\tilde{M}$ as
\be
(M^{-1})_{\mu\nu}=\sum_{\alpha\beta}Q_{\mu\alpha}
Q_{\nu\beta}(\tilde{M}^{-1})_{\alpha\beta}  ~~~.
\ee

Since $\tilde{M}$ is block diagonal the evaluation of its
determinant and inverse elements is straightforward.  We find
\be
det~\tilde{M}=D_{O}D_{E}
\ee
where $D_{O}$ is the determinant of the {\it odd} part of the
matrix given by
\be
\left (
\begin{array}{ccc}
 2(S_{0}-C_{0})& u_{1}-u_{2} &  u_{4}\\
u_{1}-u_{2}&\frac{1}{2}(S^{(4)}-C_{4}) &\frac{1}{2}u_{3}\\
u_{4} &  \frac{1}{2}u_{3}&\frac{1}{2}(S^{(2)}+C_{2})
\end{array}
\right )
\ee
so
\bea
\nonumber
2D_{O}=(S_{0}-C_{0})\left[(S^{(4)}-C_{4})(S^{(2)}+C_{2})-u_{3}^{2}\right]
\eea
\bea
-(u_{1}-u_{2})\left[(u_{1}-u_{2})(S^{(2)}+C_{2})-u_{3}u_{4}\right]
+u_{4}\left[u_{3}(u_{1}-u_{2})-u_{4}(S^{(4)}-C_{4})\right] ~~~.
\eea
$D_{E}$ is the determinant of the {\it even} part of the matrix
given by
\be
\left (
\begin{array}{ccc}
\frac{1}{2}(S_{0}+C_{0}) & u_{1}+u_{2} & u_{4}\\
u_{1}+u_{2}&2(S^{(4)}+C_{4})& 2u_{3}\\
u_{4}&2u_{3}& 2(S^{(2)}-C_{2})
\end{array}
\right )
\ee
where
\bea
\nonumber
D_{E}/2=(S_{0}+C_{0})\left[(S^{(4)}+C_{4})(S^{(2)}-C_{2})-u_{3}^{2}\right]
\eea
\bea
-(u_{1}+u_{2})\left[(u_{1}+u_{2})(S^{(2)}-C_{2})-u_{3}u_{4}\right]
+u_{4}\left[u_{3}(u_{1}+u_{2})-u_{4}(S^{(4)}+C_{4})\right]
\eea
Since it is easy to show that $det~Q =1$
we obtain
\be
det ~\tilde{M}=det ~Q ~det ~ M ~det ~Q=det ~ M =D_{E}D_{O}
\ee

The needed inverses
are given by
\be
(\tilde{M}^{-1})_{33}=\frac{(S_{0}-C_{0})(S^{(2)}+C_{2})-u_{4}^{2}}{D_{O}}
\ee
\be
(\tilde{M}^{-1})_{35}=\frac{(u_{1}-u_{2})u_{4}-(S_{0}-C_{0})u_{3}}{D_{O}}
\ee
\be
(\tilde{M}^{-1})_{55}=
\frac{(S_{0}-C_{0})(S^{(4)}-C_{4})-(u_{1}-u_{2})^{2}}{D_{O}}
\ee
\be
(\tilde{M}^{-1})_{44}=\frac{(S_{0}+C_{0})(S^{(2)}-C_{2})-u_{4}^{2}}{D_{E}}
\ee
\be
(\tilde{M}^{-1})_{46}=\frac{(u_{1}+u_{2})u_{4}-(S_{0}+C_{0})u_{3}}{D_{E}}
\ee
\be
(\tilde{M}^{-1})_{66}=
\frac{(S_{0}+C_{0})(S^{(4)}+C_{4})-(u_{1}+u_{2})^{2}}{D_{E}}
\ee
All odd-even inverse elements such as $(\tilde{M}^{-1})_{34} $
vanish and the rest of the elements follow using the
symmetry of $(\tilde{M}^{-1})$.  One can then easily extract the
inverse elements of $(M^{-1})$ :
\be
(M^{-1})_{33}=(M^{-1})_{44}=\frac{1}{4}(\tilde{M}^{-1})_{33}
+(\tilde{M}^{-1})_{44}
\ee
\be
(M^{-1})_{55}=(M^{-1})_{66}=\frac{1}{4}(\tilde{M}^{-1})_{55}
+(\tilde{M}^{-1})_{66}
\ee
\be
(M^{-1})_{34}=(\tilde{M}^{-1})_{44}-\frac{1}{4}(\tilde{M}^{-1})_{33}
\ee
\be
(M^{-1})_{35}=-(M^{-1})_{46}=\frac{1}{4}(\tilde{M}^{-1})_{35}
-(\tilde{M}^{-1})_{46}
\ee
\be
(M^{-1})_{36}=-(M^{-1})_{45}=\frac{1}{4}(\tilde{M}^{-1})_{35}
+(\tilde{M}^{-1})_{46}
\ee
\be
(M^{-1})_{56}=\frac{1}{4}(\tilde{M}^{-1})_{55}
-(\tilde{M}^{-1})_{66}
\ee

In terms of $f$ and its derivatives we find that we can write
\be
D_{E}=2\sigma^{3}\left[\kappa_{E}^{(0)}\kappa_{E}^{(2)}
-(1+f)[\kappa_{E}^{(1)}]^{2}\right]
\label{eq:A84}
\ee
\be
D_{O}=\frac{1}{2}\sigma^{3}\left[-\kappa_{O}^{(0)}\kappa_{O}^{(2)}
-(1-f)[\kappa_{O}^{(1)}]^{2}\right]
\label{eq:A85}
\ee
\be
D_{O}(\tilde{M}^{-1})_{33}=L^{2}\sigma^{2} (1-f)\kappa_{O}^{(0)}
\ee
\be
D_{O}(\tilde{M}^{-1})_{35}=L\sigma^{2} (1-f)\kappa_{O}^{(1)}
\ee
\be
D_{O}(\tilde{M}^{-1})_{55}=-\sigma^{2}\kappa_{O}^{(2)}
\ee
\be
D_{E}(\tilde{M}^{-1})_{44}=L^{2}\sigma^{2} (1+f)\kappa_{E}^{(0)}
\ee
\be
D_{E}(\tilde{M}^{-1})_{46}=L\sigma^{2} (1+f)\kappa_{E}^{(1)}
\ee
\be
D_{E}(\tilde{M}^{-1})_{66}=\sigma^{2}\kappa_{E}^{(2)}   ~~~.
\ee
These results give the explicit $L$ dependences of the various
matrix elements.  The $\kappa$'s are independent of $L$ and given
by
\be
\kappa_{E}^{(1)}=(\nabla^{2}f)'-\frac{f'(\nabla^{2}f)_{+}}{1+f}
\label{eq:A92}
\ee
\be
\kappa_{E}^{(2)}=(1+f)(\nabla^{4}f)_{+}-[(\nabla^{2}f)_{+}]^{2}
\label{eq:A93}
\ee
\be
\kappa_{E}^{(0)}=(f'')_{-}-\frac{(f')^{2}}{1+f}
\label{eq:A94}
\ee
\be
\kappa_{O}^{(1)}=(\nabla^{2}f)'+\frac{f'(\nabla^{2}f)_{-}}{1-f}
\label{eq:A95}
\ee
\be
\kappa_{O}^{(2)}=(1-f)(\nabla^{4}f)_{-}+[(\nabla^{2}f)_{-}]^{2}
\label{eq:A96}
\ee
\be
\kappa_{O}^{(0)}=-(f'')_{+}-\frac{(f')^{2}}{1-f}
\label{eq:A97}
\ee
where we have introduced the notation
\be
A_{\pm}=A(x)\pm A(0) ~~~.
\ee
%
% ACKNOWLEDGEMENTS
%
\centerline{acknowledgements}
I thank Rob Wickham for many useful conversations concerning this
material.
This work was supported in part by the MRSEC Program of the National Science
Foundation under Award Number DMR-9400379.
%
% REFERENCES
%

\begin{table}
\begin{tabular}{c|c|c|}
\hline
$\kappa_{E}^{(1)}  $ & $ 2x                         $  & $  0              $\\
$\kappa_{E}^{(2)}  $ & $ 16                         $  & $  2n             $\\
$\kappa_{E}^{(0)}  $ & $ x^{2}                      $  & $   1             $\\
$\kappa_{O}^{(0)}  $ & $ \frac{x^{4}}{12}           $  & $   0             $\\
$\kappa_{O}^{(1)}  $ & $ -\frac{x^{3}}{2}           $  & $ -2n             $\\
$\kappa_{O}^{(2)}  $ & $ -2x^{4}                    $  & $   1             $\\
$D_{E}             $ & $ 16\sigma^{3}x^{2}          $  & $ 4n\sigma^{3}    $\\
$D_{O}             $ & $ \sigma^{3}\frac{x^{8}}{48} $  & $ n\sigma^{3}     $\\
$det ~M            $ & $ \frac{\sigma^{6}x^{10}}{3} $  & $ 4n^{2}\sigma^{6}$\\
\hline
\end{tabular}
\caption{small $x$ (left) and large $x$ values for various quantities
(far left) defined in the text}
\end{table}

\begin{figure}
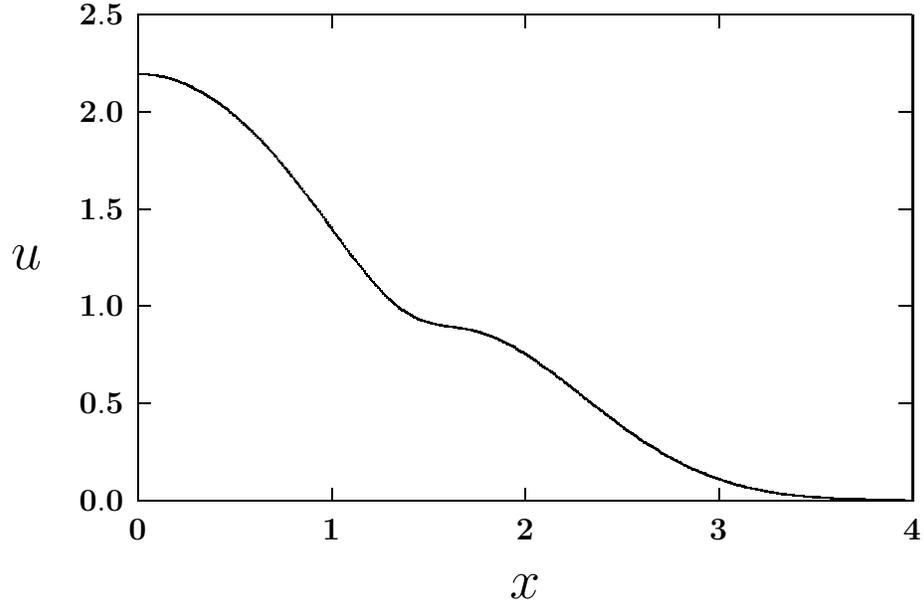

% GNUPLOT: LaTeX picture
\setlength{\unitlength}{0.240900pt}
\ifx\plotpoint\undefined\newsavebox{\plotpoint}\fi
\sbox{\plotpoint}{\rule[-0.200pt]{0.400pt}{0.400pt}}%
%
\vspace{0.5in}
\caption{The most probable scaled velocity of vortex $2$ multiplied by the
magnitude of the scaled distance of separation x, $u=\vec{x}
\cdot\vec{u}(2)L/\Gamma c$ versus x.  This velocity is directed along
$\hat{x}$ the line connecting the two tagged vortices.  The most
probable velocity of vortex 1 is equal and opposite to that of vortex
2.}
\end{figure}
\pagebreak
\begin{figure}
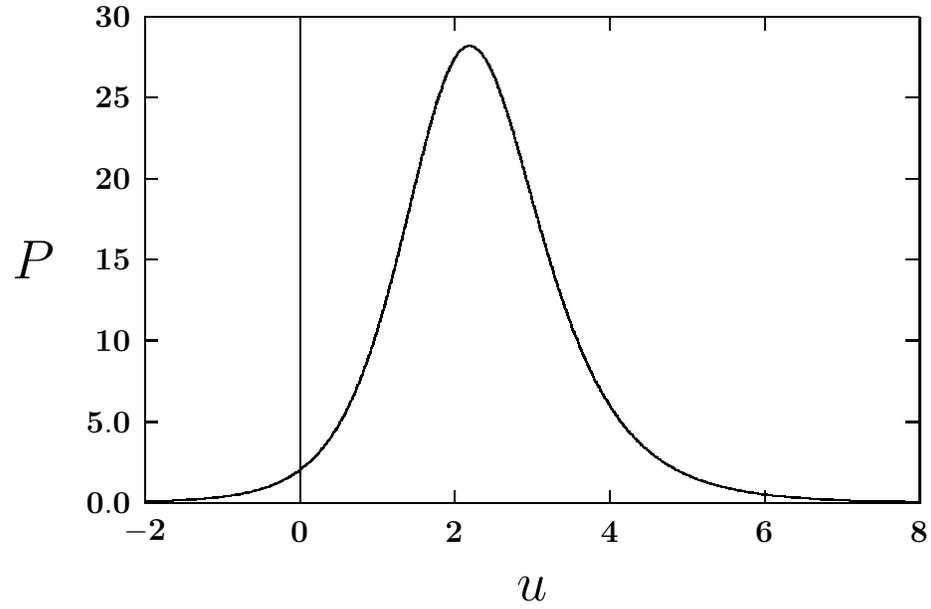

% GNUPLOT: LaTeX picture
\setlength{\unitlength}{0.240900pt}
\ifx\plotpoint\undefined\newsavebox{\plotpoint}\fi
\sbox{\plotpoint}{\rule[-0.200pt]{0.400pt}{0.400pt}}%
%
\vspace{0.5in}
\caption{Probability distribution (unnormalized) for the scaled
longitudinal velocity
$u=\vec{x}\cdot\vec{u}(2)L/\Gamma c=-\vec{x}\cdot\vec{u}(1)L/\Gamma c$
in the small scaled distance $x$ limit.  The peak in this curve gives
the small $x$ limit of the quantity plotted in Fig.1.}

\end{figure}
\pagebreak
\begin{figure}
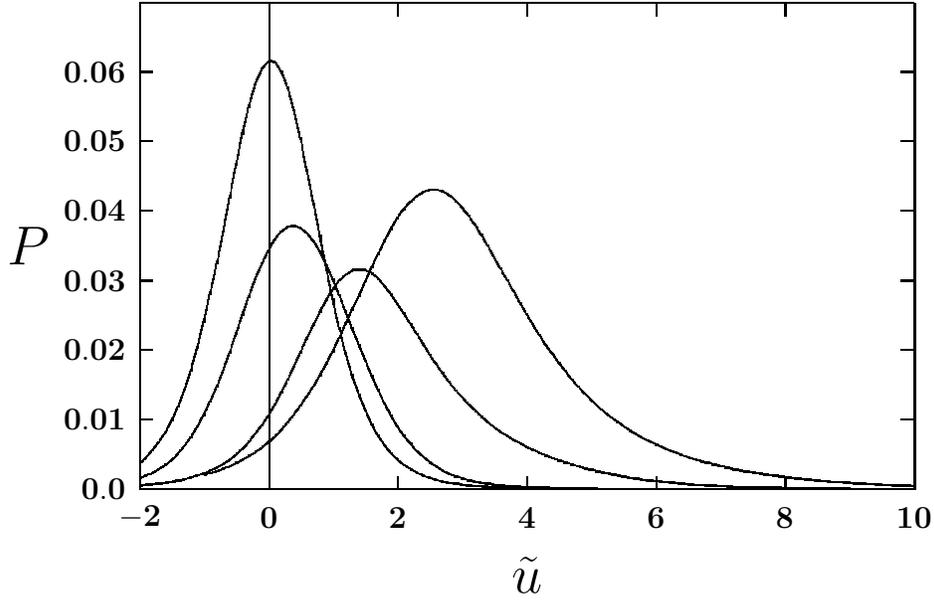

% GNUPLOT: LaTeX picture
\setlength{\unitlength}{0.240900pt}
\ifx\plotpoint\undefined\newsavebox{\plotpoint}\fi
\sbox{\plotpoint}{\rule[-0.200pt]{0.400pt}{0.400pt}}%
%
\vspace{0.5in}
\caption{Plot of unnormalized probability $P[\vec{v}_{1},\vec{v}_{2}]$
for different values of the scaled distance x between the two tagged
vortices versus the scaled velocity in the center of mass
$\tilde{u}=\hat{x}\cdot\vec{u}(2)L/\Gamma c
=-\hat{x}\cdot\vec{u}(1)L/\Gamma c$.  The normalization changes with
$x$ so the different heights of the curves is not significant in
this plot. The curves, as one moves from left to right, are
labelled by $x=3.0,2.0,1.0$ and $0.7$ respectively.}
\end{figure}
\pagebreak
\begin{figure}
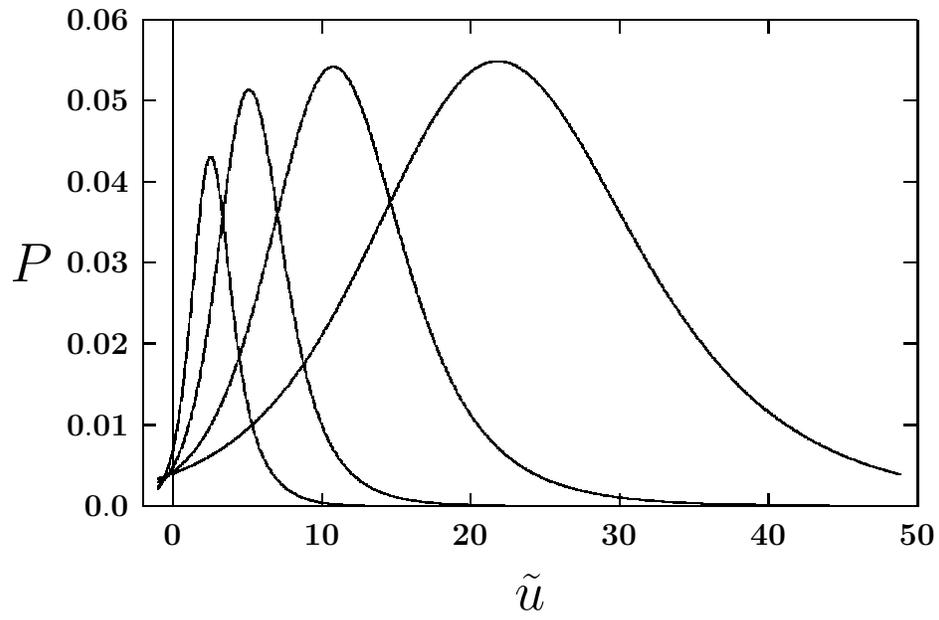

% GNUPLOT: LaTeX picture
\setlength{\unitlength}{0.240900pt}
\ifx\plotpoint\undefined\newsavebox{\plotpoint}\fi
\sbox{\plotpoint}{\rule[-0.200pt]{0.400pt}{0.400pt}}%

\vspace{0.5in}
\caption{Same as Fig.3 except $x=0.7,0.4,0.2$ and $0.1$ as one moves
from left to right.}
\end{figure}


\begin{thebibliography}{99}
\bibitem{POTENTIAL} Typically, $V$ is taken to have the form:
                    $V = \frac{1}{4} (\psi_{0}^{2} - \psi^{2})^{2}$.
\end{thebibliography}
\end{document}